\documentclass[preprint,floatfix
,aps,prd,superscriptaddress]{revtex4-2}

\usepackage[utf8]{inputenc}
\usepackage{graphicx}
\usepackage{tikz}
\usepackage{amssymb,amsmath,mathtools,comment,hyperref}
\usepackage{appendix}
\usepackage[utf8]{inputenc}
\usepackage{physics}

\usepackage{array, makecell}
\usepackage{cellspace} %
\usepackage{extarrows}
\usepackage{relsize}

\usepackage{bm}
\usepackage{mathrsfs}
\usepackage{xcolor}

\usepackage{dcolumn}
\usepackage{subcaption}

\begin{document}


\title{Entanglement gegradation in causal diamonds}

\author{H. E. Camblong}
\affiliation{Department of Physics and Astronomy, University of San Francisco, San Francisco, California 94117-1080, USA}
\author{A. Chakraborty}
\affiliation{Institute for Quantum Computing, University of Waterloo, Waterloo, Ontario N2L 3G1, Canada}
\author{P. Lopez-Duque}
\affiliation{Department of Physics, University of Houston, Houston, Texas 77024-5005, USA}

\author{C. R. Ord\'{o}\~{n}ez}
\affiliation{Department of Physics, University of Houston, Houston, Texas 77024-5005, USA}
\date{\today}

\begin{abstract}
Entanglement degradation appears to be a generic prediction in relativistic quantum information whenever horizons restrict access to a region of spacetime. This property has been previously explored in connection with the Unruh effect, where a bipartite entangled system composed of an inertial observer (Alice) and a uniformly accelerated observer (Rob) was studied, with entanglement degradation caused by the relative acceleration---and 
with equivalent results for the case when Alice is freely falling into a black hole and Rob experiences a constant proper acceleration as a stationary near-horizon observer. 
In this work, we show that a similar degradation also occurs in the case of an entangled system composed of an inertial observer (Alice) and a ``diamond observer'' (Dave) with a finite lifetime. The condition of a finite lifetime is equivalent to the restriction of Dave's access within a causal diamond. Specifically, if the system starts in a maximally entangled state, prepared from Alice's perspective, entanglement degradation is enforced by the presence of the diamond's causal horizons.
 
\end{abstract}

\maketitle
\newpage

\section{\label{sec:Introduction }Introduction\protect }

Entanglement is a purely quantum-mechanical correlation between systems that plays a central role in quantum information science~\cite{nielsen00}, where it is used as an important resource for miscellaneous tasks~\cite{qi-app-1,qi-app-2,qi-app-3} and applications in quantum computing~\cite{qc-app-1,qc-app-2,qc-app-3}.
It is generally expected that the entanglement between two systems is prone to destruction due to environmental effects---a phenomenon known as entanglement degradation, which is a form of decoherence~\cite{entanglement-death:2004}.
Hence, to keep two systems entangled, it is essential to find the sources of decoherence and quantify the amount of degradation of the quantum correlation between the systems. In this work, we focus on a source of entanglement degradation involving the relativistic decoherence effects on entanglement between two scalar field modes.

In recent years, many cutting-edge experiments have reached a limit in which relativistic effects become relevant \cite{sidhu2021advances}; therefore, a thorough understanding of entanglement in a fully relativistic framework is critically important at the experimental level. The theory of relativistic quantum information (RQI) is now a well-established field that grew from the pioneering works of Refs.~\cite{peres2004quantum,Peres2002QuantumRelativity,alsing2003teleportation,reznik2003entanglement}, and which has uncovered
novel relativistic quantum properties of entanglement, and quantum information more generally. 
For example, multiple roles played by the causal propagator in RQI were studied in Ref.~\cite{tjoa-1}, 
and a truly relativistic quantum teleportation protocol was established in~\cite{tjoa-2}, that considered relativistic propagation of a quantum field. 
In addition to theoretical consistency, such realizations would eventually allow the inclusion of relativistic effects to improve quantum tasks.
From the perspective of two inertial observers, the entanglement between the two global field modes remains unchanged, as expected \cite{Peres2002QuantumRelativity}. These ideas led to a prediction of entanglement degradation when accelerated observers are involved, for scalar fields~\cite{Fuentes-Schuller2005AliceFrames}, and also for fermionic fields~\cite{Alsing2006EntanglementFrames}; and further analyzed for the relative acceleration of a falling and a stationary observer in the Schwarzschild black hole geometry~\cite{Martin-Martinez2010UnveilingHole,Martin-Martinez2009FermionicHole}, along with a variety of related results on entanglement degradation of global field modes at the relativistic 
level~\cite{Martin-Martinez2010QuantumEntanglement,Martin-Martinez2010PopulationFrames,Martin-Martinez2011RedistributionFrames,Alsing2012Observer-dependentEntanglement,Richter2015DegradationEffect}.

The context for these RQI results is the framework of quantum field theory in curved spacetime, which was originally developed to deal with fields in the presence of gravitational backgrounds~\cite{birrell-davies}, including black holes~\cite{Hawking:1974-HE1,Hawking:1975-HE2}, and from the perspective of noninertial observers~\cite{Unruh:1976,birrell-davies}.
In quantum field theory, the notions of vacuum, particles, and horizons are deeply intertwined: the particle content is observer dependent except for the simple case of inertial reference frames~\cite{birrell-davies}. A major outcome of this program was the development of profound connections between general relativity and quantum physics, leading to black hole thermodynamics~\cite{Bardeen:1973-BHmech,Bekenstein:1973-entropy,Hawking:1976_BHTh},
and the Hawking and Unruh effects. In particular, even in flat spacetime, a constantly accelerated observer would detect a thermal distribution of particles at the Unruh-Davies temperature $T_U=a/{2\pi}$, where $a$ is the acceleration---this is the Unruh effect~\cite{Unruh:1976,birrell-davies, Takagi:1986, Crispino:2008}. (Here and throughout the paper, we adopt natural units with $c = \hbar =k_{B} =1$; and the metric convention with signature $(-,+)$ corresponding to the ordering of time and spatial coordinates.) A similar effect is Hawking radiation by black holes~\cite{Hawking:1974-HE1,Hawking:1975-HE2}, with Hawking temperature $T_H=\kappa/{2\pi}$, where $\kappa$ is the black hole's surface gravity.
One critical discovery of this program is that causally restricting an observer to a partial region of spacetime, e.g., in the presence of horizons: (i) leads to a thermal state because the observer does not have access to all the degrees of freedom of the quantum field theory; (ii) creates inequivalent vacuum states, leading to the observation of particles for some observers in the vacuum of another set of observers~\cite{birrell-davies}.
For the Hawking effect, this is enforced by the black hole's event horizon, while for the Unruh effect, there exists a Killing horizon associated with the accelerated motion.

As far as entanglement degradation is concerned, the results of Refs.~\cite{Fuentes-Schuller2005AliceFrames,Alsing2006EntanglementFrames,Martin-Martinez2010UnveilingHole,Martin-Martinez2009FermionicHole, Martin-Martinez2010QuantumEntanglement,Martin-Martinez2010PopulationFrames,Martin-Martinez2011RedistributionFrames,Alsing2012Observer-dependentEntanglement,Richter2015DegradationEffect} basically show that entanglement is generically observer dependent when considering noninertial frames, curved spacetimes, and/or the presence of horizons that restrict spacetime access.
The general framework for these studies of entanglement degradation involves a pure maximally entangled Bell state, typically given by
\begin{equation}
\ket{\Psi}
=
\frac{1}{\sqrt{2}}
\bigg(
\ket{0_{j_A}}_{\! A}
\otimes
\ket{0_{j_B}}_{\! B}
+
\ket{1_{j_A}}_{\! A}
\otimes
\ket{1_{j_B}}_{\! B} \bigg)
 \; , 
 \label{eq:psi_ab}
\end{equation}
where the subscripts $A$ and $B$ denote the two observers, Alice and Bob, who have detectors sensitive to each of the two entangled modes $j_A$ and $j_B$ of the field (involving different frequencies and directions of propagation).
In Eq.~(\ref{eq:psi_ab}), the states $ \ket{0_{j}}$ and $ \ket{1_{j}}$ are the 0-particle and 1-particle states corresponding to each mode $j$ (from the perspective of an inertial observer). Now, a noninertial observer in the bipartite system would perceive  
 the initially prepared pure state of Eq.~(\ref{eq:psi_ab}) as a mixed state. Specifically, if Bob is a uniformly accelerated observer, the states
$ \ket{0_{j_B}}_{\! B}$ and $\ket{1_{j_B}}_{\! B}$ appear as mixed---this is the key property of the Unruh effect~\cite{Unruh:1976,birrell-davies,Crispino:2008}: a uniformly accelerated observer perceives the Minkowski vacuum as a mixed, thermal state with temperature $T_U$.
Moreover, the mixed nature of the state~(\ref{eq:psi_ab}) correspondingly degrades the degree of entanglement, as shown in Refs.~\cite{Fuentes-Schuller2005AliceFrames,Alsing2006EntanglementFrames,Martin-Martinez2010UnveilingHole,Martin-Martinez2009FermionicHole, Martin-Martinez2010QuantumEntanglement,Martin-Martinez2010PopulationFrames,Martin-Martinez2011RedistributionFrames,Alsing2012Observer-dependentEntanglement,Richter2015DegradationEffect}.
 
In all the studies of entanglement degradation mentioned above, the observers have infinite lifetime. By contrast, one can consider an observer with a finite lifetime $\mathcal{T} = 2 \alpha$, whose causal access is restricted to a finite region of spacetime. This region, called the causal diamond or double cone, is the intersection between the future light cone at some initial time $t_0$ and the past light cone at some later time $t_f$.
  
From the viewpoint of a finite lifetime (diamond) observer, the boundary of the associated causal diamond is an apparent horizon similar to the Rindler horizon, and restricting causal access. It is this specific setup that we consider in our paper for the analysis of the entanglement and total correlations of a bipartite system composed of an inertial observer (Alice) and a diamond observer (Dave).
The existence of thermodynamic effects associated with a causal diamond was clearly spelled out in the pioneering work of Ref.~\cite{martinetti-rovelli:2003} within the thermal time hypothesis~\cite{Connes:1994-Thermal}, and building on earlier work on modular flows~\cite{Bisognano-Wichmann, Hislop-Longo,Haag_LQP} that showed that a conformally invariant vacuum is subject to the KMS condition~\cite{KMS_Kubo,KMS_MS}. In essence, a diamond observer can detect thermal particles in the Minkowski vacuum, as was further corroborated in a series of papers~\cite{martinetti-2, Ida-etal_2013, Su2016SpacetimeDiamonds,DeLorenzo_light-cone,jacobson2019gravitational,Chakraborty:2022oqs, Foo2020GeneratingMirror}. Additional work has found generalizations that map the diamond physics to de Sitter spacetime~\cite{Tian_DS, Good-etal_DS} and black hole horizons~\cite{DeLorenzo_light-cone,DeLorenzo_BH}.
 
The thermal behavior is governed by the diamond temperature 
 \begin{equation}
 T_D = \frac{2}{\pi \mathcal{T} } \; ,
 \label{eq:diamond-temperature}
 \end{equation}
which proportionally scales inversely with the lifetime $\mathcal{T}$ of the observer and is the generalization of the Unruh-Davies temperature for finite lifetime observers. Using a state of the form~(\ref{eq:psi_ab})---initially prepared as maximally entangled from the perspective of inertial observers---we show the emergence of entanglement degradation properties from the restriction of causal horizons that limit the diamond. This analysis clarifies the origin of observer-dependent entanglement for finite lifetime systems introduced in the recent Ref.~\cite{Foo2020GeneratingMirror}. As part of this analysis, we elucidate the conformal mapping between Rindler space and the diamond, including the subtleties needed for entanglement degradation.
 
This article is structured as follows. In Sec.~\ref{sec:diamond-geometry}, we review the geometry and the conformal mapping needed to describe the causal diamond
and we show how each of the Rindler wedges is mapped under this transformation. In Sec.~\ref{sec:Quantization-diamond}, we use modes inside and outside the causal diamond to quantize the scalar field and calculate the Bogoliubov coefficients that relate the Minkowski and diamond modes. In Sec.~\ref{sec:Entanglement-bipartite}, we set up the formalism for entanglement, by properly generalizing Eq.~(\ref{eq:psi_ab}) to derive the properties of entanglement in bipartite systems; and apply this approach to the field modes observed by an inertial observer and a diamond observer. In Sec.~\ref{sec:conclusions}, we offer further insight into the significance of these results, and outline possible future work. The appendices expand the discussion of the conformal mapping, the geometry, and the canonical quantization in causal diamonds.

\section{Causal Diamond: Geometry and Diamond Coordinates}
\label{sec:diamond-geometry}

In this section, we introduce the geometry and coordinate setup of the causal diamond $\mathrm{D}$ in a form that will prove convenient for the quantum field theory calculations that follow.

\subsection{Diamond geometry}
\label{sec:diamond-geometry_basic}

\begin{figure}[htb]
   \centering
    \includegraphics[width=0.5\linewidth]{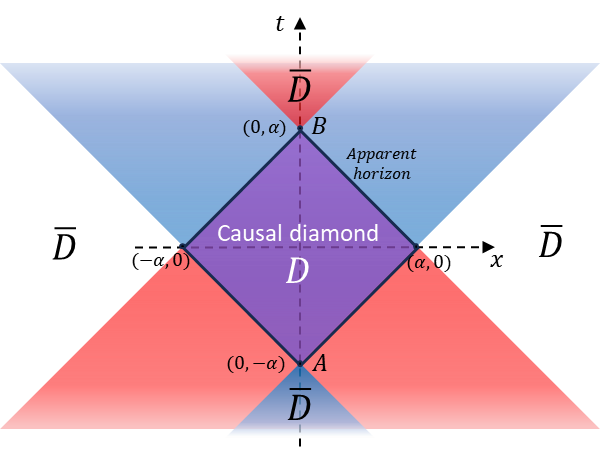}
    \caption{Causal diamond $\mathrm{D}$ (in purple) 
    associated with a starting event A and an end point event B. 
    This double cone is the intersection of the forward 
    light cone of A (causal future)
    and the past light cone of B (causal past). $\bar{D}$ is the relevant part of the outside region.}
    \label{fig:causal-diamond-schematics}
\end{figure}

The basic geometry of a causal diamond, as depicted in Fig.~\ref{fig:causal-diamond-schematics}, is defined by the double cone subtended 
by the intersection of the future and past light cones of two events A (``birth") and B (``death''). This spacetime geometry encodes the physics of an
observer with a finite lifetime $\mathcal{T} = 2 \alpha$, for whom causal access is strictly restricted within region $\mathrm{D} \coloneqq \{(x,t): |t|+|x|\leq\alpha\}$.

The standard way of parametrizing this geometry is via the physics of Rindler spacetime, considering a conformal transformation that maps the finite domain of a causal diamond $\mathrm{D}$ in Minkowski spacetime into the infinite domain of the right Rindler wedge $\mathrm{R \coloneqq\{(\tilde{x},\tilde{t}): |t|\leq \tilde{x} \text{ and }  \tilde{x}\geq 0\}}$. This is the generic approach followed in all the papers dealing with the physics of the causal diamond, including the seminal work on its thermodynamics by Martinetti and Rovelli~\cite{martinetti-rovelli:2003}. This conformal approach can be traced back to the work of Ref.~\cite{Hislop-Longo}, where the modular flow of the double cone was studied using the method of the Bisognano-Wichmann theorem~\cite{Bisognano-Wichmann, Haag_LQP} for wedge regions.

\begin{figure}[htb]
   \centering
    \includegraphics[width=0.3\linewidth]{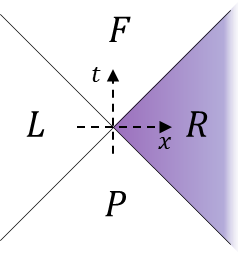}
    \caption{Rindler wedges $\mathrm{R},\mathrm{L},\mathrm{P}$, and $\mathrm{F}$. The right Rindler wedge $\mathrm{R}$
     (in purple) is mapped to a causal diamond under a conformal transformation.}
    \label{fig:rindler-wedges-schematics}
\end{figure}

The geometric transformation used in this mapping requires understanding Rindler spacetime, which is the natural framework to analyze the motion of uniformly accelerated observers. Due to their constant proper acceleration, observers are restricted to a region of Minkowski spacetime called a wedge. If we use standard Rindler coordinates, Minkowski spacetime is divided into four regions: the $\mathrm{L}$ (left), $\mathrm{R}$ (right),  $\mathrm{F}$ (future), and $\mathrm{P}$ (past) wedges, as can be seen in Fig.~\ref{fig:rindler-wedges-schematics}. The coordinate charts $(\eta,\xi) $ are restricted to one wedge at a time; thus, as usual, the maximal extension actually consists of four separate wedge-restricted charts to cover the whole of Minkowski spacetime~\cite{birrell-davies,Crispino:2008,Olson-Ralph:2011}. 
By construction, one starts with accelerated observers, whose spacetime trajectories are in either one of the two causally disconnected wedges $\mathrm{R}$ or $\mathrm{L}$. The boundaries of these regions are apparent horizons---diamond observers in region $\mathrm{D}$ do not have causal access to the information in the relevant part $\overline{\mathrm D}$ of the outside region, and viceversa for observers in $\overline{\mathrm D}$. Thus, these are two regions that are entangled for accelerated observers, and they are the ones of interest for the causal diamond in our paper~\cite{birrell-davies, Takagi:1986, Crispino:2008}. 

A key advantage of this conformal approach is that the causal structure is unaffected by this transformation. 
It is also noteworthy that, by construction of the conformal mapping, the transformation between the Rindler and diamond spacetimes is one-to-one and covers the whole unrestricted Minkowski. Thus, this conformal mapping naturally allows to define coordinates for the interior of the diamond as well as for all of its exterior regions.
Next, in Subsec.~\ref{sec:diamond-coordinates}, we construct the complete mapping of all the regions of diamond Minkowski spacetime, filling a gap in the existing literature, where only part of the mapping has been displayed. In addition, we 
precisely define the diamond coordinates needed for a description of the quantum fields.

\subsection{Diamond-Rindler conformal transformation and diamond coordinates}
\label{sec:diamond-coordinates}

In this subsection, we briefly summarize our generalized approach and the main results, and write them in a form that is tailored for the quantum-field-theory algebra that we will use for the remainder of the paper. 

Different versions of the diamond coordinate transformations have been used in the literature~\cite{martinetti-2,casini:2011,Ida-etal_2013, Su2016SpacetimeDiamonds, DeLorenzo_light-cone, jacobson2019gravitational, Foo2020GeneratingMirror, Chakraborty:2022oqs}, following 
Ref.~\cite{martinetti-rovelli:2003}.
The unifying framework that subsumes the existing transformation variants
consists of
a systematic two-step method that defines the most general diamond coordinates via conformal transformations,
with the inclusion of a generic scaling transformation $\Lambda(\lambda)$
allowed by the scaling symmetry of Rindler space.
This approach is based on a generalization of the procedure of Ref.~\cite{Chakraborty:2022oqs}.
As we show in this paper (Appendix~\ref{sec:general-conformal-mappings}),
the composite two-step transformations are
 restricted by enforcing a condition on physical scaling that yields the correct scale dimensions.
 This guarantees that the outcome of all calculations for observables 
 return the correct value of the diamond temperature and associated field-theory quantities in a scale-independent manner. 
As a result, as we further prove in Appendix~\ref{sec:general-conformal-mappings} and 
Sec.~\ref{sec:light-cone-coordinates},
the final expressions are independent of the chosen scaling $\lambda$, once the constraint is implemented. However, as the 
intermediate expressions connecting diamond and Rindler spacetime Minkowski coordinates are usually displayed in the literature, our 
unifying framework permits an easy comparison of apparently dissimilar mappings.
For example,
the case with scaling $\lambda = 2$ (which provides the maximal symmetry
between the direct and inverse conformal mappings)
has been used in most of the causal-diamond 
literature~\cite{martinetti-2,Ida-etal_2013, Su2016SpacetimeDiamonds, DeLorenzo_light-cone, Foo2020GeneratingMirror, Chakraborty:2022oqs}, 
while the case 
with  $\lambda =1$ (which has the minimalist structure with no scaling at the level of the first-step conformal transformations)
was selected in Refs.~\cite{casini:2011,jacobson2019gravitational, Chakraborty:2022oqs}.
In short, our generalized procedure consists of two following steps: 

(i) using appropriate combinations of conformal transformations~\cite{CFT_DiFranceso};

(ii) covering the Rindler wedge R with a Rindler coordinate chart $(\eta,\xi)$.

The first step involves selecting an appropriate combination of special conformal transformations $K(\rho)$, dilatation scalings $\Lambda(\lambda)$, and translations $T(\alpha)$ to generate a composite, conformal, one-to-one mapping  $(\tilde{t},\tilde{x}) \longrightarrow (t,x)$ from the right Rindler wedge $\mathrm{R}$ to the diamond region $\mathrm{D}$, described via Minkowski coordinates $(t,x)$. The appropriate composition is given by
 $T(-\alpha)\circ K(1/2\alpha)\circ \Lambda(\lambda)$, which results in
\begin{equation}
    \frac{ t }{\alpha} 
     = \frac{ 2 \, \tilde{t}/\tilde{\alpha}
    }{ (\tilde{x}/\tilde{\alpha}+1)^2-(\tilde{t}/\tilde{\alpha})^2}
        \; , \qquad  
   \frac{ x }{\alpha} 
  = - 
  \frac{
  1-(\tilde{x}/\tilde{\alpha})^2+(\tilde{t}/\tilde{\alpha})^2 
    }{  (\tilde{x}/\tilde{\alpha}+1)^2-(\tilde{t}/\tilde{\alpha})^2 }
  \; ,
  \label{eq:rindler-diamond_MR_R-to-D} 
\end{equation}
where, in these expressions and in the remainder of this section,
the parameter $\lambda$ only appears through the rescaled variable 
$\tilde{\alpha}=2\alpha/\lambda$. 
The corresponding inverse transformation is
\begin{equation}
    \frac{ \tilde{t} }{\tilde{\alpha}} 
     = \frac{ 2 \, t/\alpha
    }{ (x/\alpha-1)^2-(t/\alpha)^2}
        \; , \qquad  
   \frac{ \tilde{x} }{\tilde{\alpha}} 
  = 
  \frac{
  1-(x/\alpha)^2+(t/\alpha)^2 
    }{  (x/\alpha-1)^2-(t/\alpha)^2 }
  \; .
  \label{eq:rindler-diamond_MR_D-to-R}
\end{equation}

The second step uses an appropriate rescaled version of the standard relation between the Minkowski coordinates $(\tilde{t},\tilde{x})$ of the wedge and the Rindler coordinates $(\eta,\xi)$~\cite{birrell-davies,Takagi:1986, Crispino:2008}, i.e., a Rindler mapping $(\eta, \xi) \longrightarrow (\tilde{t},\tilde{x})$. This mapping, when the scaling factor is $\lambda$ (and $\epsilon = \pm 1$ for $\mathrm{D}$ and  $\overline{\mathrm D}$ respectively), is given by:
 \begin{equation}
   \frac{ \tilde{t} }{ \tilde{\alpha}}= \epsilon \, e^{2 \xi/\alpha}\sinh(2 \eta/\alpha)
    \; ,
    \qquad \qquad 
      \frac{\tilde{x}}{\tilde{\alpha}}  = \epsilon \, e^{ 2\xi/\alpha}\cosh(2\eta/\alpha)
    \; ,
    \label{eq:Rindler-right_Minkowski}
\end{equation}
with ranges $ \eta, \xi \in (-\infty,\infty)$. 
The proper scaling in Eq.~(\ref{eq:Rindler-right_Minkowski}), i.e., the choice of the proportionality constants (scaling of the spacetime 
coordinates with $\tilde{\alpha}$ and a numerical coefficient equal to one)
is determined in Appendix~\ref{sec:general-conformal-mappings}
by a condition that governs the correct physical dimensions. As usual, the curves $\xi = \mathrm{const}$ in the conformally mapped Rindler spacetime correspond to uniformly accelerated observers with acceleration $2 (\alpha\, e^{2\xi/\alpha})^{-1}$, where $a= 2/\alpha$ is the basic acceleration parameter for the worldline $\xi=0$. In addition, for regions  $\overline{\overline{\mathrm D}}$ (conformally mapped from the corresponding Rindler wedges $\mathrm{F}$ and $\mathrm{P}$), the roles of the spatial and temporal coordinates get reversed ($\tilde{t} \leftrightarrow \tilde{x}$)---for the coordinate patches covering these regions, formulas similar to Eq.~(\ref{eq:Rindler-right_Minkowski}) can be used with such reversals.
Thus, such extensions of Eq.~(\ref{eq:Rindler-right_Minkowski}) can be used to cover globally all regions of the maximally extended diamond spacetime.

In short, combining the two steps, the parametrization mapping 
 \begin{equation}
\underbrace{(\eta, \xi)}_{\substack{\text{Rindler coordinates} \\ \text{in Rindler spacetime}
 \\ \equiv \; \text{diamond coordinates} }}
\stackrel{\substack{\text{Rindler} \\  \text{parametrization}}}{\xleftrightarrow{\hspace*{1in}} }
\underbrace{(\tilde{t},\tilde{x}) }_{\substack{\text{Minkowski coordinates} \\ \text{in Rindler spacetime}}}
\stackrel{\substack{\text{conformal} \\  \text{transformation}}}{\xleftrightarrow{\hspace*{1in}}}
\underbrace{(t,x)}_{\substack{\text{Minkowski coordinates} \\ \text{in diamond spacetime}}}
\label{eq:mapping_diamond-Rindler-Minkowski}
\end{equation}
is established.
For our purposes, the existence of this mapping suggests the presence of entanglement degradation for observers moving inside a causal diamond $\mathrm{D}$, just as for accelerated observers in the right Rindler wedge $\mathrm{R}$. 
This mapping effectively labels the double cone with ``diamond coordinates'' $(\eta, \xi)$, as they have been called in the recent literature~\cite{Su2016SpacetimeDiamonds}.  
The final expressions for the composite mapping $(t,x) \longrightarrow (\eta, \xi)$,
and their inverses, have to be set up separately for the 4 regions; see Eq.~(\ref{eq:diamond-coordinates_MR}) and surrounding paragraph in Appendix~\ref{sec:general-conformal-mappings}.
In short, {\em the transformation $(t,x) \longrightarrow (\eta, \xi)$ is $\lambda$-independent, even though the mapping~(\ref{eq:rindler-diamond_MR_R-to-D})\/} between Minkowski coordinates of the diamond and Rindler spacetimes does depend on the scaling $\lambda$.
Finally, these results can be rewritten more elegantly in terms of the light-cone variables, as displayed in the next section,
 in a form that simplifies the algebra to follow for the remainder of the paper.

In summary, our approach generalizes the method of Ref.~\cite{Chakraborty:2022oqs} to provide a unifying framework for the conformal mappings and definition of diamond coordinates. The different variants of diamond coordinates can be realized within this framework in step (i) by uniquely fixing the values of the parameters of the special conformal transformation $K(\rho)$ and translation $T(\alpha)$ to generate a diamond of size $2 \mathcal{T}$; and by picking an arbitrary value of $\lambda$ in $\Lambda(\lambda)$. Once the composite conformal transformation is chosen, care must be taken when defining the Rindler coordinate chart in step (ii). There are specific scaling conditions that must be met so that the physical dimensions in diamond space are compatible with those in Rindler space (see Appendix~\ref{sec:general-conformal-mappings} for details). The value of the diamond temperature and associated field-theory consequences (including quantum effects such as entanglement) are dependent on these adjustments. Our generalized framework ensures that we recover the correct value of the unique diamond temperature and that we correctly account for entanglement.

\subsection{Light-cone coordinates}
\label{sec:light-cone-coordinates}

A convenient alternative to the coordinate transformation used in the previous section, with diamond coordinates~(\ref{eq:diamond-coordinates_MR}), can be restated in terms of sets of light-cone coordinates,
 \begin{equation}
 U_{\sigma} =t +\sigma x
\; , \; \; \; \; 
 \tilde{U}_{\sigma} =\tilde{t} +\sigma \tilde{x}
\; , \; \; \; \; 
u_{\sigma}= \epsilon ( \eta + \sigma \xi )
\; , 
 \label{eq:light-cone-coordinates}
 \end{equation}
where $\sigma= \pm 1$ labels the propagation direction (corresponding to left and right movers respectively). 
This set includes the advanced $U_{+} \equiv V= t +x$ and retarded
 $U_{-} \equiv U= t -x$
 Minkowski null coordinates;
 their counterparts for Rindler spacetime, 
 $\tilde{U}_{\pm} \equiv \tilde{V}, \tilde{U}= \tilde{t} \pm \tilde{x}$;
 and the null diamond, Rindler-induced coordinates
 $u_{+} \equiv v=\epsilon( \eta +\xi ) $
 and 
 $u_{-} \equiv u= \epsilon ( \eta - \xi )
$.
The sign reversal with $\epsilon = \pm 1$ for the diamond coordinates in Eq.~(\ref{eq:light-cone-coordinates}) makes the null coordinates be always future directed.
With these definitions of null coordinates, the following alternative transformation equations can be established as equivalent to Eq.~(\ref{eq:rindler-diamond_MR_D-to-R}).
First,
 \begin{equation}
\frac{\tilde{V}}{\tilde{\alpha}}
  =
  \frac{1+V/\alpha}{1-V/\alpha}
    \;  , \; \; \; \;
\frac{\tilde{U}}{\tilde{\alpha}}
=
-  \frac{1-U/\alpha}{1+U/\alpha} 
  \; ;
  \label{eq:rindler-diamond_Mink-lightcone_R-to-D}
  \end{equation}
and their inversion, equivalent to~(\ref{eq:rindler-diamond_MR_R-to-D}), simply involves switching the roles of $V$ and $U$.
Remarkably, Eq.~(\ref{eq:rindler-diamond_Mink-lightcone_R-to-D}), like the original~(\ref{eq:rindler-diamond_MR_D-to-R}), is valid in all wedges. 
Second, in terms of the Rindler-induced diamond variables $u_{\sigma}$, 
 \begin{equation}
   e^{2 v/\alpha}
  =
   \frac{1+V/\alpha}{1-V/\alpha}
    \;  , \; \; \; \;
e^{2 u/\alpha}
=
 \frac{1+U/\alpha}{1-U/\alpha} 
  \; ,
  \label{eq:rindler-diamond_lightcone1_R-to-D_int}
  \end{equation}
for the diamond interior D;
and 
   \begin{equation}
   e^{2 \bar{v}/\alpha}
  =
   \frac{V/\alpha-1}{V/\alpha+ 1}
    \;  , \; \; \; \;
e^{2 \bar{u}/\alpha}
=
 \frac{U/\alpha-1}{U/\alpha+ 1} 
  \; ,
  \label{eq:rindler-diamond_lightcone1_R-to-D_ext}
  \end{equation}
for the relevant diamond exterior $\overline{\mathrm D}$.
In Eqs.~(\ref{eq:rindler-diamond_lightcone1_R-to-D_int}) and (\ref{eq:rindler-diamond_lightcone1_R-to-D_ext}), and for the remainder of the paper,
we will denote the null diamond variables separately in the two relevant regions; specifically, $({v}, {u})$ and $(\bar{v}, \bar{u})$ refer to the diamond interior and exterior, respectively. These relations can be easily inverted to show that
\begin{equation}
\frac{V}{\alpha } = \tanh (v/\alpha)
\; \; , \; \;  \frac{U}{\alpha } = \tanh (u/\alpha) \; ,
 \label{eq:rindler-diamond_lightcone1_D-to-R_int}
\end{equation}
for the diamond interior D; and $V/\alpha =- \coth (v/\alpha)$ along with $U/\alpha  = -\coth (u/\alpha)$ for the relevant diamond exterior $\overline{\mathrm D}$.

Moreover, the basic null-coordinate transformation Eqs.~(\ref{eq:rindler-diamond_lightcone1_R-to-D_int})--(\ref{eq:rindler-diamond_lightcone1_D-to-R_int})
are independent of the scaling factor $\lambda$, in a form that is especially useful for  the study of the field modes to be addressed 
in Sec.~\ref{sec:Quantization-diamond}.
The technicalities associated with rescalings for these light-cone variables are further analyzed in Appendix~\ref{sec:general-conformal-mappings}.

\begin{figure}[htb]
    \begin{subfigure}{0.35\textwidth}
        \includegraphics[width=\linewidth]{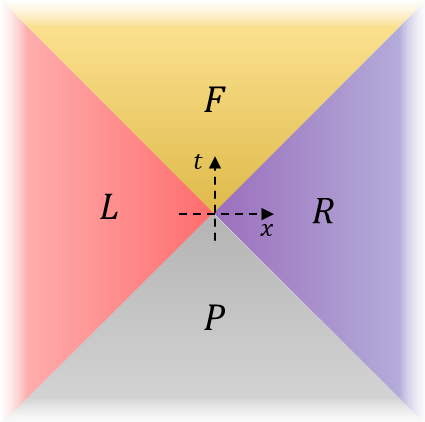}
    \end{subfigure}
    \hspace{1cm}
    \begin{subfigure}{0.35\textwidth}
        \includegraphics[width=\linewidth]{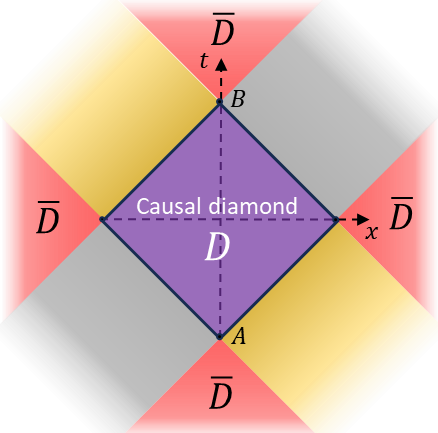}
    \end{subfigure}
    \caption{\label{fig:wedges} Transformation of the Rindler wedges under the conformal map.
The right wedge $\mathrm{R}$ in purple maps into the interior diamond $\mathrm{D}$,
the left wedge $\mathrm{L}$ in red maps into the four regions of $\overline{\mathrm D}$, and the gray and yellow past $\mathrm{P}$ and future $\mathrm{F }$ wedges map into the corresponding color-coded regions of  $\overline{\overline{\mathrm D}}$.}
\end{figure}

Finally, the global mapping, i.e., the conformal transformation defined via Eq.~(\ref{eq:rindler-diamond_MR_D-to-R}) or Eq.~(\ref{eq:rindler-diamond_Mink-lightcone_R-to-D}), is shown in Fig.~\ref{fig:wedges}, including the specific mappings of the four Rindler wedges to the diamond spacetime. 
This mapping turns the right Rindler wedge $\mathrm{R}$, $\epsilon = +1$,  into the diamond interior $\mathrm{D}$; and the left Rindler wedge $\mathrm{L}$, $\epsilon = -1$, into the relevant parts $\overline{\mathrm D}$ of the diamond exterior entangled with the interior. There are also exterior regions $\overline{\overline{\mathrm D}}$ that can be reached by analytic continuation from $\mathrm{D}$ and $\overline{\mathrm D}$, which correspond to the  regions $\mathrm{F}$ and $\mathrm{P}$ of Rindler spacetime, as shown in Fig.~\ref{fig:wedges}. The details are discussed in Appendix~\ref{sec:diamond-geometry_global}.

\section{Field Quantization of the Causal Diamond}
\label{sec:Quantization-diamond}

For our analysis of the entanglement properties associated with quantum fields in a causal diamond, we will consider a free, minimally coupled, scalar massless field $\Phi$ in  $(1+1)$-dimensional Minkowski spacetime. This simple model, despite exhibiting well-known infrared divergences~\cite{Coleman:1973}, is a good laboratory to display the basic quantization procedure and the existence of inequivalent vacua and thermal properties for the Unruh effect and generalizations.
Quantization in higher dimensions works in basically the same manner, and, for most generic quantum properties, extensions to more general fields 
are straightforward in principle.~\cite{birrell-davies, Takagi:1986, Crispino:2008}.

\subsection{Field modes in Minkowski and diamond coordinates}

We start our analysis by finding an orthonormal set of field modes adapted to each coordinate system, and follow the standard approach~\cite{birrell-davies, Takagi:1986, Crispino:2008} to relate the associated canonical quantizations of the scalar field $\Phi$. For the casual diamond, this involves a comparison between the quantizations in Minkowski coordinates and diamond coordinates via the mapping of Sec.~\ref{sec:diamond-geometry}.

In Minkowski coordinates, the modes of the massless, minimally coupled scalar field satisfy the Klein-Gordon wave equation
\begin{equation}
(\partial_t^2-\partial_x^2)
\Phi=\partial_V\partial_U \Phi=0  \; .
\label{eq:K-G-eq_Minkowski}
\end{equation}
Thus, they involve a complete set of normalized positive frequency eigenfunctions of the Killing vector $\partial_t$, including left- and right-moving Minkowski modes~\cite{Crispino:2008}
\begin{equation}
f_{+,k} (V) =  ( 4 \pi k)^{-1/2} \, e^{-ikV}
\; , \; \; 
 f_{-,k} (U) =  ( 4 \pi k)^{-1/2} \, e^{-ikU}
 \; ,
 \label{eq:K-G-modes_Minkowski}
\end{equation}
respectively. As usual, we simplify Eqs.~(\ref{eq:K-G-eq_Minkowski}) and (\ref{eq:K-G-modes_Minkowski}) by the use of the Minkowski null coordinates $U=t-x$ and $V=t+x$ [cf. Eq.~(\ref{eq:light-cone-coordinates})].

Similarly, in diamond coordinates  $(\eta, \xi)$, the Klein-Gordon equation reads
\begin{equation}
(\partial_\eta^2-\partial_\xi^2)=\partial_v\partial_u \Phi= 0  \; ,
\label{eq:K-G-eq_diamond}
\end{equation}
with the diamond null coordinates $v = \epsilon (\eta+ \xi)$ and $u = \epsilon (\eta- \xi)$ [from Eq.~(\ref{eq:light-cone-coordinates})], where we label the two different regions with  $\epsilon= + \equiv \mathrm{int}$ for the diamond interior D and  $\epsilon= - \equiv \mathrm{ext}$ for the relevant part of diamond exterior $\overline{\mathrm D}$. This labeling originates with Eq.~(\ref{eq:Rindler-right_Minkowski}); as pointed out in Sec.~\ref{sec:diamond-coordinates}, these definitions are extended to all regions of the global diamond spacetime using four separate coordinate patches. In these coordinates, Eq.~(\ref{eq:K-G-eq_diamond}) takes basically the same form as Eq.~(\ref{eq:K-G-eq_Minkowski}) due to conformal invariance; this can also be established directly from the metric (Appendix~\ref{sec:general-conformal-mappings}).
The corresponding normalized positive frequency eigenfunctions of the Killing vector $\partial_{\epsilon  \eta}$ are generally
 $\displaystyle g_{\sigma, \omega} (u_{\sigma})=\frac{1}{\sqrt{4\pi\omega}}
 e^{-i\omega u_{\sigma}}$, 
where the variables $u_{\sigma}$ are chosen separately in each region of diamond spacetime and $\omega$ stands for the diamond frequencies.
Then, {\em the diamond modes\/}, restated in terms of the Minkowski null variables, Eq.~(\ref{eq:light-cone-coordinates}), are defined with separate support in each region D and $\overline{\mathrm D}$ according to
   \begin{equation}
    g_{\sigma, \omega}^{(\epsilon)}(U_{\sigma} )=
 \frac{1}{\sqrt{4\pi\omega}}
 e^{-i\omega u_{\sigma} (U_{\sigma}) } 
\,
\theta (\epsilon (\alpha -|U_{\sigma}|))
\; ,
 \label{eq:K-G-modes-generic_diamond}
\end{equation} 
where $\theta(z)$ is the Heaviside theta function.  Explicitly, using Eqs.~(\ref{eq:rindler-diamond_lightcone1_R-to-D_int}) and (\ref{eq:rindler-diamond_lightcone1_R-to-D_ext}), the left- and right-moving diamond modes, with support in the interior region D, become 
  \begin{equation}
 g_{+, \omega}^{(\mathrm{int})}(V)
 =
 \frac{1}{\sqrt{4\pi\omega}}
 e^{-i\omega v (V)} \,
\theta  (\alpha -|V|))
 =
  \frac{1}{\sqrt{4\pi\omega}}
 \left(\frac{1+V/\alpha}{1-V/\alpha}\right)^{-i\omega \alpha /2}
 \theta (\alpha-|V|)
\; ,
\label{eq:int-plus-mode}
\end{equation}
and
\begin{equation}
  g_{-, \omega}^{(\mathrm{int})}(U)
  =
  \frac{1}{\sqrt{4\pi\omega}}
 e^{-i\omega u (U)} 
 \,
\theta (\alpha -|U|)
 =
 \frac{1}{\sqrt{4\pi\omega}}
 \left(\frac{1+U/\alpha}{1-U/\alpha}\right)^{-i\omega \alpha /2}
 \theta (\alpha-|U|) 
\; .
\label{eq:int-minus-mode}
\end{equation}
Similarly, the left- and right-moving diamond modes with support in the exterior region $\overline{\mathrm D}$ are
\begin{equation}
 g_{+, \omega}^{(\mathrm{ext})}(V)
 =
 \frac{1}{\sqrt{4\pi\omega}}
 e^{-i\omega \bar{v} (V)} 
  \,
\theta ( |V|-\alpha)
 = 
 \frac{1}{\sqrt{4\pi\omega}}
 \left( \frac{V/\alpha+1}{V/\alpha-1} \right)^{i\omega \alpha /2}
 \theta (|V|-\alpha)
 \; ,
\label{eq:ext-plus-mode}
\end{equation}
and
\begin{equation}
  g_{-, \omega}^{(\mathrm{ext})}(U)
  =
  \frac{1}{\sqrt{4\pi\omega}}
 e^{-i\omega \bar{u} (U)} 
  = 
  \frac{1}{\sqrt{4\pi\omega}}
 \left( \frac{U/\alpha+1}{U/\alpha-1} \right)^{i\omega \alpha /2}
 \theta (|U|-\alpha)
 \; .
\label{eq:ext-minus-mode}
\end{equation}

The quantization of the field can be carried out by following the canonical procedure in Minkowski and diamond coordinates. This is reviewed in Appendix~\ref{sec:appCanonQuant_BogCoefs} along the calculation of the Bogoliubov coefficients that transform between Minkowski modes and diamond modes. In the next subsection,
 we generalize the analytic continuation technique originally developed by Unruh for Rindler spacetime~\cite{Unruh:1976}, which
 generates ``Unruh-diamond modes'' by an extension of the diamond modes that only includes positive frequencies. 
 This provides specific information about the vacuum that is critical for the thermal interpretation and entanglement properties of 
 causal diamonds.

\subsection{Unruh-diamond modes and the relation between the vacua}
\label{sec:Unruh-diamond-modes}

The Bogoliubov coefficients provide a systematic way of relating the different vacuum states
(see Appendix~\ref{sec:appCanonQuant_BogCoefs}).
 It is possible to consider a simplified form of these transformations via the construction of linear combinations of diamond modes that are Minkowski positive frequency. Effectively, such modes model the Minkowski vacuum and permit an efficient way of writing the general expressions needed for questions of relativistic quantum information. This technique consists in the use of analytical continuations of the diamond modes with complex-plane properties (in the variables $U_{\sigma}$) that guarantees their positive frequency of nature with respect to the Minkowski vacuum. 

Let us first consider the modes in region D inside the causal diamond, as given in Eqs.~(\ref{eq:int-plus-mode}) and (\ref{eq:int-minus-mode}); 
and those in region $\overline{ \mathrm{D} }$ outside the causal diamond, as given in Eqs.~(\ref{eq:ext-plus-mode}) and (\ref{eq:ext-minus-mode}).
Each one of these can be analytically continued to the complementary region by using the same approach pioneered by Unruh for acceleration radiation in Rindler spacetime~\cite{Unruh:1976}.

As is the case of Rindler spacetime, the justification for this method is based on the fact that, when covering a limited region of the diamond spacetime with a coordinate patch, we have to perform a mandatory analytic continuation if the solution were to be valid globally.
However, because of the functional form involved, Eqs.~(\ref{eq:int-plus-mode})--(\ref{eq:ext-minus-mode}), these are multivalued functions and a specific choice of branch needs to be made. In order to relate the diamond modes with the global Minkowski modes, what is needed is the particular analytic continuation that only consists of positive frequencies. 
Then, the positive frequency analytic continuations of the internal (starting with region D) and external modes  (starting with region  $\overline{ \mathrm{D} }$) are the Unruh-diamond modes,
\begin{equation}
\begin{aligned}
h_{\sigma, \omega}^{(\mathrm{int})} (U_{\sigma} )
& = 
\left(
1 -  e^{-\pi \omega \alpha } \right)^{-1/2}
\, 
\bigl[
 g_{\sigma, \omega}^{(\mathrm{int})} (U_{\sigma} )
+ e^{-\pi \omega \alpha/2 } 
\,
g_{\sigma , \omega}^{(\mathrm{ext})* } (U_{\sigma} )
\bigr]
\\
& = \frac{1}{\sqrt{2 \sinh (\pi \omega \alpha/2 )}}
\, 
\bigl[
 e^{\pi \omega \alpha/4 } 
 \,
 g_{\sigma, \omega}^{ (\mathrm{int}) } (U_{\sigma} )
+ e^{-\pi \omega \alpha/4 } 
\,
g_{\sigma, \omega}^{ (\mathrm{ext})* } (U_{\sigma} )
\bigr]
\; ,
\label{eq:Unruh-diamond_modes_int}
\end{aligned}
\end{equation}
and
\begin{equation}
\begin{aligned}
h_{\sigma, \omega}^{ (\mathrm{ext}) } (U_{\sigma} )
& = 
\left( 1 -  e^{-\pi \omega \alpha } \right)^{-1/2}
\, 
\bigl[
 g_{\sigma, \omega}^{ (\mathrm{ext})} (U_{\sigma} )
+ e^{-\pi \omega \alpha/2 } 
\,
g_{\sigma, \omega}^{ (\mathrm{int})* } (U_{\sigma} )
\bigr]
\\
& = \frac{1}{\sqrt{2 \sinh (\pi \omega \alpha/2 )}}
\,
\bigl[  
 e^{\pi \omega \alpha/4 } 
 \,
 g_{\sigma, \omega}^{ (\mathrm{ext}) } (U_{\sigma} )
+ e^{-\pi \omega \alpha/4 } 
\,
g_{\sigma, \omega}^{(\mathrm{int})* } (U_{\sigma} )
\bigr] 
\; .
\label{eq:Unruh-diamond_modes_ext}
\end{aligned}
\end{equation}

In what follows, as is common in the literature of relativistic quantum information~\cite{Fuentes-Schuller2005AliceFrames,Martin-Martinez2010UnveilingHole,Crispino:2008}, we will parametrize these linear combinations, which model the Minkowski vacuum and are inequivalent 
to the original diamond modes, via the variable
\begin{equation}
 r_{\omega} = \tanh^{-1} \! \left( e^{-\pi \alpha \omega/2} \right)
 \label{eq:r-parameter}
\; .
\end{equation}
Then,
\begin{equation}
\begin{aligned}
h_{\sigma, \omega}^{ (\mathrm{int}) }
 = 
 \cosh r_{\omega} 
 \,
 g_{\sigma, \omega}^{ (\mathrm{int}) }
+
\sinh r_{\omega} 
\,
g_{\sigma, \omega}^{(\mathrm{ext})* }
\\
h_{\sigma, \omega}^{ (\mathrm{ext}) }
= 
\cosh r_{\omega} 
 \,
 g_{\sigma, \omega}^{ (\mathrm{ext}) }
+ 
\sinh r_{\omega} 
\,
g_{\sigma, \omega}^{ (\mathrm{int})* }
\; .
\label{eq:Unruh-diamond_modes_r-parametrized}
\end{aligned}
\end{equation}
The transformation of field operators can be directly obtained by inversion of the modes~(\ref{eq:Unruh-diamond_modes_r-parametrized}) to give the same generic field operator $\Phi$; then,
 \begin{equation}
\begin{aligned}
c_{\sigma, \omega}^{ (\mathrm{int}) }
 = 
 \cosh r_{\omega} 
 \,
 b_{\sigma, \omega}^{ (\mathrm{int}) }
-
\sinh r_{\omega} 
\,
b_{\sigma, \omega}^{(\mathrm{ext})\dagger }
\\
c_{\sigma, \omega}^{ (\mathrm{ext}) }
= 
\cosh r_{\omega} 
 \,
 b_{\sigma, \omega}^{ (\mathrm{ext}) }
-
\sinh r_{\omega} 
\,
b_{\sigma, \omega}^{ (\mathrm{int}) \dagger}
\; .
\label{eq:Unruh-diamond_operators_r-parametrized}
\end{aligned}
\end{equation}
These linear combinations are effective Bogoliubov transformations between the diamond modes and the appropriate linear combinations of Minkowski positive-frequency modes $h_{\sigma, \omega}^{ (\mathrm{int}) }$ and  $h_{\sigma, \omega}^{ (\mathrm{ext}) }$.
With the given effective Bogoliubov coefficients, 
$\alpha^{ (\mathrm{eff}) }_{\omega}
= \cosh r_{\omega}
=
[ 2 \sinh (\pi \omega \alpha/2 ) ]^{-1/2} \,
 e^{\pi \omega \alpha/4 }
$ and
$\beta^{ (\mathrm{eff}) }_{\omega}
= \sinh r_{\omega} =
[ 2 \sinh (\pi \omega \alpha/2 ) ]^{-1/2} \,
 e^{-\pi \omega \alpha/4 }$, a straightforward formal argument shows that the coefficient ratio 
$\beta^{ (\mathrm{eff}) }_{\omega}/\alpha^{ (\mathrm{eff}) }_{\omega} 
= \sinh r_{\omega}/\cosh r_{\omega}=\tanh r_{\omega}=e^{-\pi \alpha} $ 
yields a Boltzmann factor $(\tanh r_{\omega})^2 = |\beta^{ (\mathrm{eff}) }_{\omega}/\alpha^{ (\mathrm{eff}) }_{\omega}|^2 = e^{- \beta \omega}$ for a thermal state. For example, this can be shown rigorously by writing the Unruh-diamond vacuum $ \ket{0}^{\! \mathcal{U}}$, which satisfies
\begin{equation}
c^{(\mathrm{int})}_{\pm,\omega} \, 
\left| 0 \right\rangle^{\!\mathcal U} = 0
\; \; \; \; 
\mathrm{and}
\; \; \; \; 
c^{(\mathrm{ext})}_{\pm,\omega} \, 
\left| 0 \right\rangle^{\!\mathcal U} = 0
\label{eq:Unruh-diamond-vacuum}
\end{equation}
for all diamond frequencies $\omega$, in terms of the diamond vacuum $ \ket{0}^{\! \mathcal{D}}$ of Eq.~(\ref{eq:diamond-vacuum}); this is a straightforward consequence of field-operator relations~(\ref{eq:Unruh-diamond_operators_r-parametrized}), which imply~\cite{Wald_QFT-curved, Milburn_Qoptics}
\begin{align}
 \ket{0}^{\!\mathcal{U}}
&
= Z^{-1/2}
\prod_{\sigma,\omega}
\exp \left(
e^{-\pi \alpha \omega/2}
\sum_{\sigma = \pm}
b_{\sigma, \omega}^{ (\mathrm{int}) \dagger}
b_{\sigma, \omega}^{ (\mathrm{ext}) \dagger}
\right)
\,
 \ket{0}^{\! \mathcal{D}}
\label{eq:Unruh-vacuum_alpha-parametrized_1}
\\
&
=
Z^{-1/2}
\prod_{\sigma, \omega}
\sum_{n=0}^{\infty}
e^{-n \pi \alpha \omega/2}
\ket{n_{\sigma, \omega}}^{\! (\mathrm{int})}
\otimes
\ket{n_{\sigma, \omega}}^{\! (\mathrm{ext})} 
\label{eq:Unruh-vacuum_alpha-parametrized_2}
\; ,
\end{align}
where $Z= \left( 1-e^{-\pi\omega\alpha} \right)^{1/2}$. Then, the corresponding vacuum reduced density matrix $\rho_{\mathrm D} $ of the interior diamond region $\mathrm{D}$ can be derived by tracing out the exterior degrees of freedom (of region $\overline{\mathrm D}$); the result is
\begin{equation}
\rho_{\mathrm{D}}
=
\operatorname{Tr}_{_{\overline{\mathrm D}}}
\,
\bigl(
\ket{0}^{\! \mathcal{U} }
\prescript{\mathcal{U} \! \! }{}{\bra{0}}
\bigr)
=
Z^{-1} e^{-\pi \alpha H}
\; ,
\label{eq:diamond-thermal-density-matrix}
 \end{equation}
where 
$H = \sum_{\sigma = \pm} 
\int d \omega \, \omega \,
b_{\sigma, \omega}^{ (\mathrm{int}) \dagger} b_{\sigma, \omega}^{ (\mathrm{int})}
$,
represents a mixed state of thermal nature with an inverse temperature parameter $\beta = \pi \alpha $ that  yields the diamond temperature, Eq.~(\ref{eq:diamond-temperature}).

It should be noted that Eq.~(\ref{eq:Unruh-vacuum_alpha-parametrized_2}), leading to a thermal density matrix~(\ref{eq:diamond-thermal-density-matrix}),
is a two-mode squeezed state of the field, involving only two non vacuum modes. Moreover, it has the simple product structure,
\begin{equation}
 \ket{0}^{\!\mathcal{U}}
 =
\bigotimes_{j}
 \ket{0_{j} }^{\!\mathcal{U}}
 \; ,
\end{equation}
where the modes are labeled by the multi-index $j  = (\sigma, \omega)$, and, as in Eq.~(\ref{eq:Unruh-vacuum_alpha-parametrized_2}),
\begin{equation}
 \ket{0_{\sigma, \omega} }^{\! \mathcal{U}}
=
\frac{1}{\cosh  r_{\omega} }
\,
\sum_{n=0}^{\infty}
\tanh^{n}  r_{\omega} 
\,
\ket{n_{\sigma, \omega}}^{\! (\mathrm{int})}
\otimes
\ket{n_{\sigma, \omega}}^{\! (\mathrm{ext})}
\label{eq:Unruh-vacuum_r-parametrized}
\; .
\end{equation}
Then, from the Unruh-diamond vacuum state~(\ref{eq:Unruh-vacuum_r-parametrized}), applying the creation operators, the one-particle Unruh-diamond states are
$ \ket{1_{\sigma, \omega}}^{\! \mathcal{U}}
= 
c_{\sigma, \omega}^{ (\mathrm{int})
 \dagger}
 \ket{0_{\sigma, \omega} }^{\! \mathcal{U}}$,
 whence
 \begin{equation} 
 \ket{1_{\sigma, \omega}}^{\! \mathcal{U}}
=
\frac{1}{\cosh^2 r_{\omega} }
\,
\sum_{n=0}^{\infty}
\tanh^n r_{\omega}  \sqrt{(n+1)}
\ket{(n+1)_{\sigma, \omega}}^{\! (\mathrm{int})}
\otimes\ket{n_{\sigma, \omega}}^{\! (\mathrm{ext})} 
\; ,
\label{eq:Unruh-one-particle_r-parametrized}
 \end{equation} 
and so on. Equations~(\ref{eq:Unruh-vacuum_r-parametrized}) and (\ref{eq:Unruh-one-particle_r-parametrized}) spell out the thermal and two-squeezed nature of the vacuum, and are key ingredients for the description of the state used for entanglement degradation in the next section.

\section{Entanglement in causal diamonds: relativistic bipartite systems}
\label{sec:Entanglement-bipartite}

In this section, we consider the standard setup for the analysis of entanglement in different frames. In this approach, a scalar field is initially prepared in a state
\begin{equation}
\frac{1}{\sqrt{2}}
\, 
\left(
\ket{0_{j}}^{\! \mathcal{M}}
\! \otimes
\ket{0_{j'}}^{\! \mathcal{M}}
+
\ket{ 1_{j}}^{\! \mathcal{M}}
\! \otimes
\ket{ 1_{j'} }^{\! \mathcal{M}}
\right) 
\; ,
\label{eq:Minkowski_max-entangled}
\end{equation}
which is maximally entangled from the perspective of any two inertial observers. This is a two-mode entangled state of the Bell-state form in Eq.~\ref{eq:psi_ab}). In this initial field configuration, described in the inertial Minkowski basis $ \mathcal{M}$, all the modes except two,  labeled by the multi-indices $j=(\sigma, k)$ and $j'=(\sigma', k')$, are in the vacuum state. The states $ \ket{ 0_{j}}^{\! \mathcal{M} }$ and $\ket{ 1_{j}}^{\! \mathcal{M}}$ are the Minkowski vacuum and one-particle excited states of the mode labeled by $j$. Here, the global Minkowski-Fock vacuum state, defined by the absence of particle excitations of all the modes, is the tensor product
 $\ket{0}^{\! \mathcal{M}}
 =
\bigotimes_{\sigma, k}
 \ket{0_{\sigma, k} }^{\! \mathcal{M}}
$. 

We can consider the traditional operational procedure in which the modes are probed by one inertial observer, Alice (A), with a detector sensitive only to mode $j$; and another observer, Bob (B), with a detector sensitive only to mode $j'$. In this manner, we can focus only on the modes $j$ and $j'$, formally tracing out all the other vacuum modes. For a noninteracting field, these modes do not mix; thus, the reduced state~(\ref{eq:Minkowski_max-entangled}) labeled by $j$ and $j'$ is a pure state. In the analysis below, we will only consider the physics generated by such reduced states. 
    
\subsection{Entanglement description in causal diamonds: States and density matrices}
    
Our goal is to find the degree of entanglement when a state of the form~(\ref{eq:Minkowski_max-entangled}) is described by diamond observers. The essence of the effect can be analyzed in the simplest setting, when the first observer, Alice (A), remains inertial, and another observer, Dave (D), is restricted to a causal diamond. We can then replace the nontrivial two-mode entangled Bell state~(\ref{eq:Minkowski_max-entangled}) by the corresponding state
\begin{equation}
\ket{\Psi}
=
\frac{1}{\sqrt{2}}
\, 
\left(
\ket{0_{j}}^{\! \mathcal{M}}
\! \otimes
\ket{0_{j'}}^{\! \mathcal{U}}
+
\ket{ 1_{j}}^{\! \mathcal{M}}
\! \otimes
\ket{ 1_{j'} }^{\! \mathcal{U}}
\right) 
%
\; ,
\label{eq:psi_Alice-Dave}
\end{equation}
where the Unruh-diamond basis ${\mathcal{U}} $ is used for the second observer, Dave. Correspondingly, the Dave modes
 $j'=(\sigma, \omega')$ involve
 the diamond frequencies $\omega'$.
 The key point is that the Minkowski and Unruh-diamond bases share the same vacuum
 $\ket{0}^{\! \mathcal{M}} =  \ket{0}^{\! \mathcal{U}}
 =
\bigotimes_{\sigma, \omega}
 \ket{0_{\sigma, \omega} }^{\! \mathcal{U}}
$, which is different from the proper diamond vacuum.
This is due to the analytic-continuation structure of the Unruh-diamond modes via Eqs.~(\ref{eq:Unruh-diamond_modes_int}) and (\ref{eq:Unruh-diamond_modes_ext}). This analytic structure guarantees that, while the Minkowski modes are a continuous infinite superposition of diamond modes,  
the transformation from the Unruh-diamond basis $\mathcal{U}$ to the proper diamond basis $\mathcal{D}$ only involves one mode at a time,
but including both the interior and exterior regions. Thus, the transformation between the states with respect to $\mathcal{U}$ and $\mathcal{D}$ can be derived from the Bogoliubov transformation~(\ref{eq:Unruh-diamond_operators_r-parametrized}), giving Eqs.~(\ref{eq:Unruh-vacuum_alpha-parametrized_2}) and
(\ref{eq:Unruh-vacuum_r-parametrized}) for the Unruh-diamond vacuum, and (\ref{eq:Unruh-one-particle_r-parametrized}) for the one-particle states.
This assignment is subtle, and the replacement is justified for scalar fields with the single-mode approximation for uniformly accelerated observers, as discussed in Ref.~\cite{Bruschi-et-al_Unruh-single-mode}. Here, we will use a similar approach for diamond observers, for which the conformal mapping preserves the basic setup.

In terms of the canonical observers defined above, the states $\ket{0}_{A} \equiv \ket{0}^{\!\mathcal M}$ and $\ket{0}_{D} \equiv \ket{0}^{\!\mathcal U}$ 
are the vacuum states for Alice and Dave respectively. In the state $\ket{\Psi}$, defined by Eq.~(\ref{eq:psi_Alice-Dave}), two arbitrary inertial observers would observe maximal correlations, but these correlations are altered (degraded) when perceived by a diamond observer.
More generally, this description captures the structure of a bipartite system in which one of the observers cannot access the information in a certain region of spacetime. In effect, even though we set up the state with two observers, the second observer, Dave, corresponding to region $\mathrm{D}$, is entangled with region $\overline{\mathrm D}$, according to Eqs.~(\ref{eq:Unruh-vacuum_r-parametrized}) and (\ref{eq:Unruh-one-particle_r-parametrized}).
Using the same notations for the observers and regions should pose no ambiguity, and highlight their one-to-one correspondence, but we will label the
observers with italics ($D$ and $\overline{D}$) and the regions with roman characters ($\mathrm{D}$ and $\overline{\mathrm D}$).
In conclusion, for the physical description of the setup defined by Eq.~(\ref{eq:psi_Alice-Dave}), the density matrix can be written down for the total system formed by the inertial observer, Alice; the diamond observer, Dave; and a third, hypothetical observer in the exterior diamond region, AntiDave. 
Thus, the states in the Minkowski/Unruh-diamond basis are states of a tripartite system $(A,D,\overline{D})$.
From the tripartite system, one can define the bipartition Alice-Dave $(A,D)$ and, similarly, the bipartition Alice-AntiDave $(A,\overline{D})$.
The subtle changes in the nature of entanglement arise from the restriction of information available to one observer at a time, in each of the bipartitions, as we describe next.

The tripartite density matrix 
\begin{equation}
\rho_{AD \overline{D}}=
\ket{\Psi}
\!
\bra{\Psi}
\; 
\label{eq:tripartite-density-matrix} 
\end{equation}
represents the total state of the system. From the viewpoint of the observers Alice and Dave, i.e., for the bipartite system  $(A,D)$, a partial trace is needed to exclude the degrees of freedom of AntiDave. This procedure is formally the same as the one involved in the derivation of Eq.~(\ref{eq:diamond-thermal-density-matrix}), but now for the composite system that also includes Alice. Clearly, as in the results described in Eq.~(\ref{eq:diamond-thermal-density-matrix}), the outcome is a mixed thermal state, which both encodes the thermal nature of the diamond and leads to an alteration of the entanglement properties. Specifically, tracing out the total density matrix~(\ref{eq:tripartite-density-matrix}),
with respect to the exterior diamond degrees of freedom of AntiDave, leads to the Alice-Dave reduced density matrix
\begin{equation} 
\rho_{AD}
=
\operatorname{Tr}_{_{\overline{\mathrm D}}}
\,
\bigl(
\rho_{AD \overline{D}}
\bigr)
=
\operatorname{Tr}_{_{\overline{\mathrm D}}}
\,
\bigl(
\ket{\Psi}
\!
\bra{\Psi}
\bigr)
\; ,
\label{eq:density-matrix_AD} 
\end{equation} 
which can be expanded in terms of the tensor product of Alice's and Dave's states. This can be done as shown below, using the shorthand notation 
$ \ket{m_j,n_{j'}}
= \ket{m_j}^{\! \mathcal{M}}
\otimes
\ket{n_{j'}}^{\! \mathrm{(int)}}$ 
for the Alice-Dave states, involving only the tensor products of Minkowski and interior-diamond regions. This assumes modes $j=(\sigma, k)$ and $j'=(\sigma, \omega')$ for a particular propagation direction $\sigma$ and frequency $k$ (with respect to the basis $\mathcal M$) and $\omega'$ [with respect to the interior diamond basis $\mathcal{D}^{ \mathrm{(int)}} \equiv \mathrm{(int)}$]. Then, the Alice-Dave reduced density matrix $\rho_{AD}$ is given by
\begin{equation} 
\rho_{AD}
=
\frac{1}{2  \cosh^2 {r_{j'}} }
\, 
\sum_{n=0}^{\infty}
\tanh^{2n} {r_{j'}} 
\,
\varrho_{AD}^{(n)}
\; ,
\label{eq:density-matrix_AD-explicit} 
\end{equation} 
where 
\begin{equation} 
\begin{aligned}
\varrho_{AD}^{(n)}
     =
     \ket{0_{j},n_{j'}} 
     \bra{0_{j},n_{j'}}
& +
\frac{\sqrt{(n+1)}}{\cosh {r_{\omega'}} }
\,
\biggl[
\ket{0_{j},n_{j'}}
\bra{1_{j},(n+1)_{j'}} 
%
+ \ket{1_{j},(n+1)_{j'}}
\bra{0_{j},n_{j'}}
\biggr]
\\ 
&
+
\frac{(n+1)}{\cosh^2 {r_{\omega'}} }
\ket{1_{j},(n+1)_{j'}}
\bra{1_{j},(n+1)_{j'}}
\; .
\label{eq:density-matrix_AD-rho-n} 
\end{aligned}
\end{equation} 
Of course, a similar treatment could be carried out for the bipartite Alice-AntiDave subsystem, though this is not directly useful for the description of the physics of a finite lifetime observer.

Finally, the reduced density matrices $\rho_A$ and $\rho_D$ for the individual systems $A$ and $D$ (which will be used in the next subsection)
can be derived by tracing out the complementary states in $\rho_{AD}$ (Minkowski and diamond interior respectively): $\rho_{D} = \operatorname{Tr}_{A} \left( \rho_{AD} \right)$ and $\rho_{A} = \operatorname{Tr}_{D} \left( \rho_{AD} \right)$ [and in the case of  $\rho_A$, most easily by using the initial state~(\ref{eq:psi_Alice-Dave}) or density~(\ref{eq:tripartite-density-matrix}), and tracing out the states associated with the diamond-Unruh basis].
As a result, 
 \begin{equation} 
  \rho_{A}
 =
 \frac{1}{2}
 \left(
 \ket{0_{j}}^{\! \mathcal{M}}
\prescript{\mathcal{M} \! \! }{}{\bra{0{_j}}}
 +
  \ket{1_{j}}^{\! \mathcal{M}}
\prescript{\mathcal{M} \! \! }{}{\bra{1_{j}}}
\right) 
 \; ,
 \label{eq:rho_A} 
\end{equation} 
where the 0- and 1-particle states correspond to the Minkowski mode $j=(\sigma, k)$; and
\begin{equation} 
 \rho_{D}
 =
 \frac{1}{2  \cosh^2 {r_{j'}} }
\, 
\sum_{n=0}^{\infty}
\tanh^{2n} {r_{j'}} 
\,
\left(
1
+ \frac{n}{\sinh^2 r_{j'} }
\right)
\ket{ n_{j'} }^{\! \mathrm{(int)}} 
\prescript{\mathrm{(int)} \! \! }{}{\bra{ n_{j'} }}
\; ,
\label{eq:rho_D} 
\end{equation} 
with $n$-particle states 
corresponding to the diamond mode $j'=(\sigma, \omega')$.

\subsection{Quantum information entanglement measures in causal diamonds}

We are now ready to analyze the relevant quantum entanglement between an inertial and a diamond observer, i.e., for the bipartite Alice-Dave subsystem.
For this purpose, we will assume that the two modes $j$ and $j'$ in Eqs.~(\ref{eq:density-matrix_AD-rho-n})--(\ref{eq:rho_D}) are specified as above. Then, to simplify the notation, we will remove the mode labels in all the equations---and it is understood that the parameter $ r \equiv r_{\omega'}$ only involves one diamond frequency $\omega'$.

In principle, a complete understanding of the physics of the Alice-Dave subsystem is contained in the reduced density matrix~(\ref{eq:density-matrix_AD-explicit})--(\ref{eq:density-matrix_AD-rho-n}). However, the derivation of specific measures that quantify this entanglement is not straightforward.
Indeed, there is a large body of literature addressing this fundamental problem in quantum information theory~\cite{nielsen00,QI-theory1,QI-theory2,QI-theory3,QI-theory4,QI-theory5}, centered on entanglement and correlation measures---see details below including specific probes. In what follows, three standard techniques are used in this context, among the many other probes available~\cite{entanglement_Plenio-Virmani}:
(i) the Peres-Horodecki criterion, 
(ii) logarithmic negativity,
and (iii) quantum mutual information.
Specifically, this treatment is similar to the analyses of entanglement degradation for accelerated observers~\cite{Fuentes-Schuller2005AliceFrames},
more general accelerated motions~\cite{Mann-Villalba_entanglement-degr-motions}, and in the presence of black hole horizons~\cite{Martin-Martinez2010UnveilingHole}.

\subsubsection{Partial transpose and Peres– Horodecki criterion}

The Peres– Horodecki or PPT (positive partial transpose) criterion~\cite{Peres_partial-transpose,Horodecki_partial-transpose} provides the simplest test for some aspects of the entanglement properties of a system. It gives a necessary condition for the joint density matrix of two quantum mechanical systems $A$ and $D$ to be separable: that all the eigenvalues of its partial transpose $\rho_{AD}^{T_{A}}$ be non-negative. The partial transpose  (with respect to subsystem $A$)~\cite{Peres_partial-transpose,Horodecki_partial-transpose,Paulsen_partial-transpose} of a matrix $\rho_{AB} $ can be defined in the tensor product Hilbert space of a bipartite system $AD$ by the matrix elements 
$\bra{n^{(A)}, m^{(D)} } \rho_{AD}^{T_{A}} \ket{k^{(A)}, j^{(D)} }
=
\bra{k^{(A)}, m^{(D)} } \rho_{AD} \ket{n^{(A)}, j^{(D)} }
$.
Physically, the partial transpose $\rho_{AD}^{T_{A}}$ corresponds to exchanging the qubits for system $A$. Thus, 
(i) if at least one eigenvalue of the partial transpose is negative, then the density matrix is entangled;
and 
(ii) when all the eigenvalues are nonnegative, there is no distillable entanglement, but bound or nondistillable entanglement may still exist.

For our bipartite Alice-Dave subsystem, the operator $\rho_{AD}^{T_{A}}$ can be computed from Eqs.~(\ref{eq:density-matrix_AD-explicit}) and (\ref{eq:density-matrix_AD-explicit}), which shows that it is represented by an infinite-dimensional matrix. However, its form is greatly simplified by reducing it to block form.  Simplifying the notation with  $\gamma_{n} = \sqrt{n+1}/\cosh r$ and$    \ket{ m_{j}^{(A) } , n_{j'}^{(D)} } = \ket{m, n} $, the reduced matrix elements relating orders $(n,n+1)$ with respect to subsystem D read
$\rho^{(n) T_{A}}
=
  \ket{0, n}  \bra{0, n} 
+
\gamma_{n} \left(   \ket{0, n}  \bra{1, n+1}  +  \ket{1, n+1}  \bra{0, n} \right)
+
\gamma_{n}^2   \ket{1, n+1}  \bra{1, n+1} 
$. 
As this sequence mixes terms of different orders, a reordering of the series can be performed such that all the terms involving orders $(n,n+1)$
are hierarchically rewritten in an infinite series of the form
\begin{equation} 
\rho_{AD}^{T_{A}}
=
\frac{1}{2  \cosh^2 r }
\, 
\left[ 
\ket{0, 0}  \bra{0, 0} 
+
\sum_{n=0}^{\infty}
\tanh^{2n} r  
\,
{\mathcal R}^{(n)}
\right]
\; ,
\label{eq:density-matrix_PT-AD-explicit} 
\end{equation} 
where the reduced matrix elements ${\mathcal R}^{(n)}$ of order $(n,n+1)$ are $2 \times 2$ blocks corresponding to the subspace $\left\{  \ket{1,n}, \ket{0,n+1} \right\}$, with
\begin{equation} 
\begin{aligned}
{\mathcal R}^{(n)}
  =
  &
  \;
  \gamma_{n-1}^2  \tanh^{-2}{r}
     \ket{1, n}  \bra{1, n}  
  +
  \,
  \gamma_{n} \bigl(    \ket{1, n}  \bra{0, n+1} +  \ket{0, n+1}  \bra{1, n}  \bigr)  
\\
& 
+
\tanh^{2}{r}  \ket{0, n+1}  \bra{0, n+1} 
\; .
\label{eq:density-matrix_PT-AD-rho-n} 
\end{aligned}
\end{equation} 
From Eq.~(\ref{eq:density-matrix_PT-AD-rho-n}), the eigenvalues can be straightforwardly computed in pairs
\begin{equation} 
 \lambda_{\pm}^{(n)}
 =
 \frac{\tanh^{2n}{r}
 }{4\cosh^2{r}}
 \left(
 \frac{n}{\sinh^2{r}}
 +
 \tanh^2{r}
 \pm
 \sqrt{Z_{n}}
 \right)
   \; ,
\label{eq:density-matrix_block-eigenvalues}
 \end{equation}
 where
 \begin{equation} 
 Z_{n}
 =
 \left(
 \frac{n}{\sinh^2{r}}
 +
 \tanh^2{r}
 \right)^2
 +
 \frac{4}{\cosh^2{r}}
 \; .
\label{eq:density-matrix_block-eigenvalues_Zn}
 \end{equation} 
Equation~(\ref{eq:density-matrix_PT-AD-explicit}) shows that there is also a first eigenvalue $\lambda_{0} = 1/2 \cosh^2 r$ corresponding to the ground state. The form of Eqs.~(\ref{eq:density-matrix_block-eigenvalues})--(\ref{eq:density-matrix_block-eigenvalues_Zn}) implies that there is a negative eigenvalue $ \lambda_{-}^{(n)}$ for any finite value of the variable $r$. As $r$ is defined via the half-size $\alpha= \mathcal{T}/2$ or lifetime $\mathcal{T}$ of the causal diamond by Eq.~(\ref{eq:r-parameter}), there is distillable entanglement for any finite size of the diamond.
[This setup corresponds to the acceleration $a$ in Rindler spacetime via $\alpha = 2/a$ in the coordinate chart defined by Eq.~(\ref{eq:Rindler-right_Minkowski})]. Furthermore, from Eq.~(\ref{eq:r-parameter}), the parameter $r$ is a decreasing function of $\alpha$ and $\mathcal{T}$, with limiting values $r=0$ and $r=\infty$. Now, the negative value of $ \lambda_{-}^{(n)}$ is reduced as $r$ increases; and in the limit of zero lifetime, as $r \rightarrow \infty$, $ \lambda_{-}^{(n)}$ asymptotically approaches zero, yielding no distillable entanglement.

\subsubsection{Entanglement measures  and logarithmic negativity}

While the PPT criterion does give some insight into the entanglement properties of a composite system, what is really needed is a specific entanglement measure. Such quantity is best established axiomatically via the concept of an entanglement monotone~\cite{entanglement-monotone_Vidal}, $E (\rho)$, which is defined to be the following: 
(i) a convex mapping from density matrices to nonnegative real numbers, which 
(ii) does not increase on average under local operations and classical communication~\cite{LOCC_Bennett-et-al}.
One commonly used measure is provided by the logarithmic negativity $E_{N} (\rho)$~\cite{log-negativity_Vidal-Werner, log-negativity_Plenio}. For a bipartite state, the logarithmic negativity is defined by
\begin{equation}
\mathcal{N}_{AD}
\equiv E_{N} (\rho_{AD})
=
\log_2
 \left\Vert
  \rho_{AD}^{T_{A}} 
 \right\Vert_1 \; ,
\label{eq:log-neg_def}
\end{equation}
 where $\rho_{AD}^{T_{A}} $ is the partial transpose and $\left\Vert M \right\Vert _{1} \equiv \operatorname{Tr}  \left[  \sqrt{M^{\dag}M}  \right]$
is the trace norm of the matrix $M$~\cite{Reed-Simon}, i.e., the sum of the absolute value of its eigenvalues, so that $\mathcal{N}_{AD} = \log_2 \left( \sum_{j} |\lambda_{j} | \right)$ (summed over all the eigenstates). The logarithmic negativity satisfies a number of desirable properties, both required by the physics and computationally efficient. First and most importantly, it is an entanglement monotone~\cite{log-negativity_Vidal-Werner, log-negativity_Plenio}, as defined above. Second, it is an additive measure, unlike the ordinary negativity 
$N_{AD} =  
\left(
\left\Vert
  \rho_{AD}^{T_{A}} 
 \right\Vert_1 - 1 \right)/2$.
Third, it provides an upper bound on the so-called distillable entanglement $E_{D} (\rho_{AD})$~\cite{log-negativity_Vidal-Werner, Horodecki_2000};
in particular, when $\mathcal{N}_{AD}=0$, there is no distillable entanglement.
And finally, it has additional convenient features, including that it also provides an upper bound to teleportation capacity.

\begin{figure}[htb]
   \centering
    \includegraphics[width=0.6\linewidth]{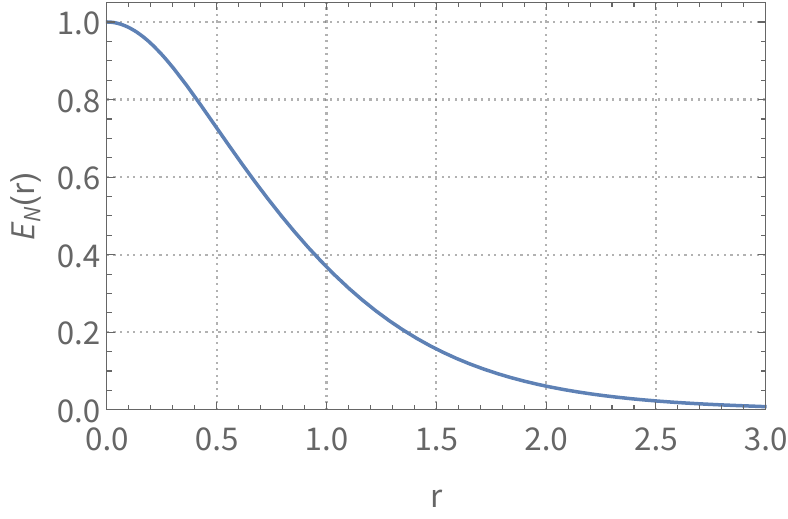}
    \caption{Logarithmic negativity as a function of $r$, showing a monotonic degradation of entanglement as the lifetime $\mathcal{T} = 2 \alpha$  is reduced.}
        \label{fig:CDs_Logarithmic_neg}
\end{figure}

For our bipartite Alice-Dave subsystem, from Eqs.~(\ref{eq:density-matrix_PT-AD-explicit})--(\ref{eq:density-matrix_block-eigenvalues_Zn}),
the relevant sum of the absolute-value eigenvalues has the form 
$
\displaystyle
\left\Vert
  \rho_{AD}^{T_{A}} 
 \right\Vert_1
=
 \lambda_{0} + \sum_{n, \pm} |\lambda^{(n)}_{\pm} |$,
so that  the logarithmic negativity becomes
\begin{equation} 
 \mathcal{N}_{AD}
 =
 \log_2
 \left(
 \frac{1}{2\cosh^2{r}}
 +
\Sigma
\right)
 \; ,
\label{eq:log-neg_CD-0}
 \end{equation} 
where
\begin{equation} 
 \Sigma 
=
 \sum_{n=0}^{\infty}
 \frac{\tanh^{2n} r}{2\cosh^2 {r} }
 \sqrt{
 \left(\frac{n}{
 \sinh^2 {r}}
 +
 \tanh^2{r}
 \right)^2
 +
 \frac{4}{\cosh^2 {r}}
 }
 \; .
 \label{eq:log-neg_CD-1}
 \end{equation} 
The logarithmic negativity function $ \mathcal{N}_{AD} (r)$ given by Eqs.~(\ref{eq:log-neg_CD-0}) and (\ref{eq:log-neg_CD-1})
is plotted in Fig.~\ref{fig:CDs_Logarithmic_neg}.
It is {\em monotonically decreasing\/} with respect to $r$, as can be easily seen from its functional form. It starts with a maximum $ \mathcal{N}_{AD}(0) = 1$ for  $\alpha= \mathcal{T}/2 = \infty$ (an unlimited lifetime), as it is to be expected for an unbound system in Minkowski spacetime.
Specifically, the unit value is required by a maximally entangled state of the form~(\ref{eq:Minkowski_max-entangled})---a pure state of maximally entangled individual states of systems A and D, initially prepared as such from the perspective of inertial observers.
Then, $ \mathcal{N}_{AD} (r)$ exhibits a gradual reduction towards the value $\displaystyle \lim_{r \rightarrow \infty} \mathcal{N}_{AD}  = 0$ for 
the opposite limit of a vanishing diamond, $\alpha =0$, i.e., zero lifetime, which most dramatically displays the entanglement degradation property.
In effect, this vanishing logarithmic negativity yields no distillable entanglement due to the degradation of quantum correlations.
These results are formally identical to the corresponding conclusions for uniformly accelerated observers (via $a=2/\alpha$).

In conclusion, the behavior of the logarithmic negativity $ \mathcal{N}_{AD} (r)$ definitively shows the {\em emergence of entanglement degradation 
that increases inversely with the lifetime or diamond size\/}---and this is associated with the restriction of causal horizons that limit the diamond.

\subsubsection{Von Neumann entropy and mutual information}

A standard measure of correlations in a system can be established in terms of information theory measures through entropy functions. At the quantum level, this is implemented with the von Neumann entropy $S=-\operatorname{Tr} \left[ \rho\log_2(\rho) \right]$, where the base-two logarithm is commonly used in quantum information theory~\cite{nielsen00}.

For a pure state of a bipartite system (composite system with two subsystems A and B), the individual entropies $S_A$ and $S_B$ can be used to measure the entanglement between their individual states---this is the well-known entanglement entropy (with $S_A=S_B$), which provides the relevant parameters for its optimal use under local operations and classical communication~\cite{LOCC_Bennett-et-al}.
However, the entanglement entropy fails to be applicable for mixed states with the same desirable properties. Instead, a preferred measure to assess the correlations of a mixed state is the  quantum mutual information 
$I(\rho_{AB}) \equiv I_{AB}$~\cite{Mutual-information_1,Mutual-information_2,Mutual-information_3,Mutual-information_4,Mutual-information_5}.
This is defined as the entropy difference $I(\rho_{AB}) \equiv I_{AB} = S_A+S_B-S_{AB}$, which is also the same as the relative entropy between the combined state $\rho_{AB}$ and the associated  pure tensor-product state, i.e., $I(\rho_{AB})  = S(\rho_{AB} \parallel \rho_A \otimes \rho_B ) $. 
Thus, $I(\rho_{AB}) $ is a nonnegative quantity that gives a distance measure of the state $\rho_{AB}$ away from being a product state. Then, it conceptually measures the amount of information that subsystem A has about subsystem B  (and reciprocally), thereby describing, for the combined system AB,
the behavior of all classical and quantum correlations.~\cite{Mutual-information_4} (A mixed state of $AB$ does include classical correlations, unlike the entanglement-only correlations of a pure state.)

\begin{figure}[htb]
   \centering
    \includegraphics[width=0.6\linewidth]{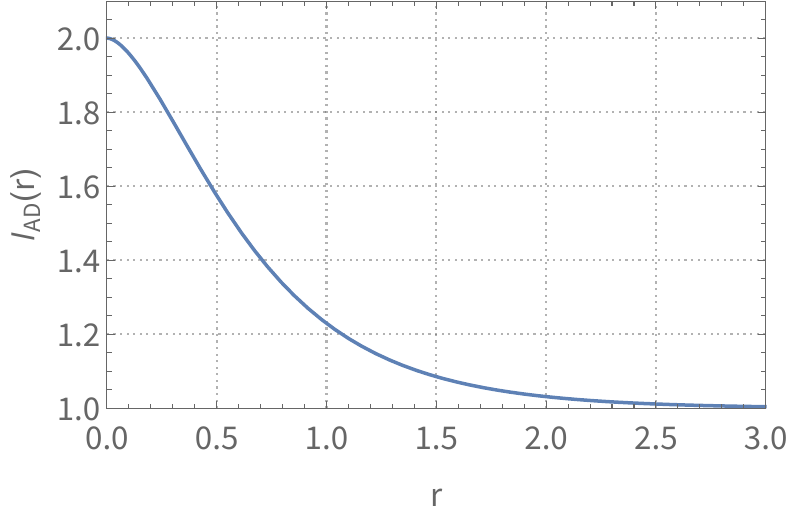}
    \caption{Mutual information as a function of $r$, showing a monotonic degradation of entanglement as the lifetime $\mathcal{T} = 2 \alpha$  is reduced.}
    \label{fig:CDs_mutual-info}
\end{figure}

We now proceed with the analysis of the mutual information for the Alice-Dave subsystem. From Eqs.~(\ref{eq:density-matrix_AD-explicit})--(\ref{eq:rho_D}),
the relevant von Neumann entropies are as follows. For the combined Alice-Dave subsystem, the joint entropy is
\begin{equation} 
 S_{AD}
 =
 -
 \frac{1}{2  \cosh^2 r }
\, 
\sum_{n=0}^{\infty}
\tanh^{2n} { r } 
\,
 \left(
 1+\frac{n+1}{\cosh^2{r}}
 \right)
 \log_2
 \left[
 \frac{
 \tanh^{2n} r }{ 2\cosh^2{r} }
 \,
 \left(
 1+\frac{n+1}{\cosh^2{r}}
 \right)
 \right] 
 \; ;
 \label{eq:S_AD}
 \end{equation} 
for the subsystem of the inertial observer, Alice, the individual entropy is 
\begin{equation} 
S_A
= 1
\; ;
 \label{eq:S_A}
\end{equation} 
and for the subsystem of the diamond observer, Dave, the individual entropy is
\begin{equation} 
S_D
=
-
\frac{1}{2  \cosh^2 r }
\, 
\sum_{n=0}^{\infty}
\tanh^{2n} r 
\,
\left(
1+
\frac{n}{\sinh^2{r}}
\right)
\log_2
\left[
\frac{\tanh^{2n} r }{ 2\cosh^2{r} }
\left(
1
+\frac{n}{\sinh^2{r}}
\right)
\right]
\; .
 \label{eq:S_D}
\end{equation} 
Therefore, the mutual information for the combined subsystem $AD$ is given by
\begin{equation} 
 I_{AD}
 =
 S_A+S_D-S_{AD}
 =1
 -
 \frac{1}{2}
 \log_2 
 \left( 
 \tanh^2{r} 
 \right)
 -
\frac{1}{2  \cosh^2 r }
\, 
\sum_{n=0}^{\infty}
\tanh^{2n} r 
 \,
 \mathcal{I}_{AD}^{(n)} 
 \; ,
 \label{eq:mutual-info}
\end{equation} 
where
\begin{equation} 
 \mathcal{I}_{AD}^{(n)} 
 =
 \left(
 1+
 \frac{n}{\sinh^2{r}}
 \right)
 \log_2 
 \left( 1+\frac{n}{\sinh^2{r}} \right)
 -
 \left(
 1+\frac{n+1}{\cosh^2{r}}
 \right)
 \log_2
 \left(
 1+\frac{n+1}{\cosh^2{r}}
 \right) 
\; .
\label{eq:mutual-info_In}
\end{equation} 
The mutual information $\mathcal{I}_{AD}$ given by Eqs.~(\ref{eq:mutual-info})--(\ref{eq:mutual-info_In}) is plotted in Fig.~\ref{fig:CDs_mutual-info}. It can be used to assess the correlations, including information about entanglement. From the strong subadditivity entropy property, with Eq.~(\ref{eq:S_A}), it is required to satisfy the bound $I_{AD} \leq 2$. Most importantly, and similarly to the logarithmic negativity~(\ref{eq:log-neg_CD-0})--(\ref{eq:log-neg_CD-1}), the function $\mathcal{I}_{AD}^{(n)}$ is monotonically decreasing with respect to the parameter $r$, and makes a smooth transition for the system $AD$ between the following:
(i) a bipartite pure and maximally entangled state ($r=0$, with $S_{AD}=0$,  $S_{A}=1$, $S_{D}=1$)
and
(ii) a bipartite entangled and maximally mixed state ($r=\infty$, with $S_{AD}=1$,  $S_{A}=1$, $S_{D}=1$).
These limiting cases are predictable: when $r=0$, this corresponds to infinite lifetime (i.e., unrestricted Minkowski spacetime, or diamond of size
$\alpha = \infty$, with zero temperature), and the mutual information is at the maximum value $I_{AD} =2$; and in the opposite limiting case, when $r \rightarrow \infty$, corresponding to zero lifetime (i.e., a diamond of zero size, $\alpha = 0$, with infinite temperature), all the subsystems under consideration are maximally mixed, yielding a mutual information  $I_{AD} =1$.
The latter limit of the bipartite system $AD$ gives a state with zero distillable entanglement, where only bound entanglement and purely classical correlations are left. These trends can be understood in terms of the concept of distributed entanglement~\cite{distributed-entaglement};
for the combined system $AD\overline{D}$, the trade-off of entanglement by pairs reduces to $ I_{AD} = 2-I_{A \overline{D}} = 1+S_{D} - S_{\overline{D}}$
(as $S_{AD}= S_{\overline{D}}$ due to the initial state being pure),  and the ensuing entanglement degradation of $AD$ from $r=0$ to $r=\infty$ is related to the entanglement sharing property~\cite{entaglement-sharing} of the global tripartite system.

In summary, we have described various measures of entanglement for the diamond system as detected by a finite lifetime observer. These results show, via the functions $\mathcal{N}_{AD}$ and $S_{AD}$,
how {\em the degradation of entanglement increases monotonically as the lifetime is reduced from infinity to zero, corresponding to the transition from an ordinary Minkowski spacetime to a system behaving thermally with infinite temperature.\/}

\section{Conclusions and Outlook}
\label{sec:conclusions}
In this paper, we generalized the approach to define diamond coordinates, which extends existing work in the literature. Moreover, we have derived the entanglement degradation perceived by finite lifetime observers, described by their restricted access to a causal diamond. In addition to elucidating a consistent realization of the conformal mapping used to describe the causal diamond, we have showed the various technical arguments that display its fundamental thermal character. In turn, thermality has a direct impact upon the diamond's entanglement, with the most important result of the present work: this quantum resource is subject to degradation. 

While our derivation shows that, for massless scalar fields, there is reduction of the maximal entanglement initially perceived by inertial observers, additional calculations would help provide robust predictions for this generic phenomenon. Possible generalizations, in progress, are:
the addition of mass to the scalar field, the corresponding behavior of fermionic fields~\cite{Alsing2006EntanglementFrames,Martin-Martinez2009FermionicHole}, and a broader analysis with wave packets and the validity of the single-mode approximation~\cite{Bruschi-et-al_Unruh-single-mode}.
Moreover, it would be useful to know how entanglement would be affected if finite lifetime were combined with uniform acceleration (Rindler)~\cite{Fuentes-Schuller2005AliceFrames}, arbitrary accelerated motions~\cite{Mann-Villalba_entanglement-degr-motions}, and black hole horizons~\cite{Martin-Martinez2010UnveilingHole}. These are different manifestations of the observer-dependent nature of relativistic entanglement~\cite{Alsing2012Observer-dependentEntanglement}. It is remarkable that the entanglement resources are perceived differently in the presence of horizons, generating a frame-dependent outcome. Clearly, as it has been suggested, a relativistically covariant notion of entanglement is needed, along with a more general notion of quantum correlations~\cite{Alsing2012Observer-dependentEntanglement}, and this would be a natural framework for all quantum-information questions about causal diamonds. In addition to its conceptual value for the completeness and consistency of the relativistic theory, such framework would yield practical results for the realization of quantum computational protocols involving observers in arbitrary configurations, motions, and restrictions in spacetime.

The thermal behavior of causal diamonds has been studied previously, but a complete understanding of the connection between thermal effects, horizons, and quantum effects still remains elusive. A tunneling approach~\cite{Parikh-Wilczek_tunneling, Banerjee-Majhi_tunneling,Vanzo-et-al_tunneling} and path integral treatments of conformal quantum mechanics (e.g., see~\cite{CQM-PI_2022}) for the analysis of time evolution within a causal diamond may be relevant for many gain insights on these connection and to generalizations mentioned in the paragraph above. It would also be worthwhile to study the effect of superpositions of causal diamonds (e.g., see~\cite{foo-superpos1,foo-superpos2}) in the entanglement of scalar field modes. We are currently exploring connections between instabilities and thermal effects in causal diamonds, which will be reported elsewhere. 

Finally, experimental results on finite lifetime observers could lead to useful tests of this relativistic quantum information framework. Some estimates of the thermal aspects of causal diamonds, interpreted with energy-scaled detectors, and involving time-dependent Stark or Zeeman effects have been mentioned in the literature~\cite{Su2016SpacetimeDiamonds}, but a more thorough investigation of the experimental setups is in order.

\section{Acknowledgements}
This material is based upon work supported by the Air Force Office of Scientific Research under Grant No. FA9550-21-1-0017 (C.R.O., A.C., and P.L.D.).
C.R.O. was partially supported by the Army Research Office (ARO), grant W911NF-23-1-0202.
H.E.C. acknowledges support by the University of San Francisco Faculty Development Fund.



\appendix

\section{Diamond geometry---General framework for conformal mappings and diamond coordinates}
 \label{sec:general-conformal-mappings}

This appendix provides a generalization of the procedure of Ref.~\cite{Chakraborty:2022oqs}, including supporting background and calculations for Sec.~\ref{sec:diamond-geometry}, with an outline of some subtleties and the generalized formulas.

The composite conformal mapping between Rindler and Minkowski coordinates for the causal diamond~(\ref{eq:mapping_diamond-Rindler-Minkowski}) consists of the two steps spelled out in Sec.~\ref{sec:diamond-geometry}:

(i) using appropriate combinations of conformal transformations~\cite{CFT_DiFranceso};

(ii) covering the Rindler wedge R with a Rindler coordinate chart $(\eta,\xi)$.

The main ingredients for step (i) are special conformal transformations $K(\rho)$, dilatations $\Lambda(\lambda)$, and translations $T(\alpha)$. The dilatation transformation $\Lambda (\lambda)$ rescales the Rindler coordinates $(\tilde{t},\tilde{x})$ 
to $x'^{\mu} = \lambda \tilde{x}^{\mu}$; at the level of the metric (see below), this implies a quadratic rescaling 
with $\lambda^2$.
 The special conformal transformation $K(\rho)$ with parameter $\rho =1/\alpha$
  (with the notation of Ref.~\cite{Chakraborty:2022oqs}) 
  brings the infinite Rindler wedge to a finite diamond of size $2 \alpha$
centered at $(0, 1/2\rho)$ in the $(t,x)$ plane. This is given by the standard definition~\cite{CFT_DiFranceso}
  \begin{equation}
  \! \! \!  \! \! \!  \! \! \!  \! \! \!
(t',x')
\overset{\displaystyle K(\rho) }{\xrightarrow{\hspace*{0.8cm}} }  
(t'',x'')
 : \; \; \; \;\; \; \; \;
 x''^\mu = \frac{x'^\mu - b^\mu (x'\cdot x')}{1-2 (b\cdot x') + (b\cdot b) (x\cdot x')} 
 \label{eq:sct_MR}
    \end{equation}
 with $b^\mu = (0,-\rho)$ being the special-conformal vector, here restricted to $(1+1)$-dimensional spacetime;
 the generalization to higher dimensions can be implemented 
 trivially~\cite{Su2016SpacetimeDiamonds,Foo2020GeneratingMirror}. 
 Translations $T(\alpha)$ are defined via $T(\alpha)x = x+\alpha$.
  The Minkowski inner product is performed
 with signature $(-,+)$ for the ordering of time and spatial coordinates.

\subsection{First step: Conformal mapping}
For the first step, we will use a conformal mapping given by the composition $M(\alpha; \lambda) = T(-\alpha) \circ K\left( 1/2\alpha \right) \circ \Lambda (\lambda)$.  This involves first a dilatation transformation $\Lambda (\lambda)$, i.e., a rescaling of the Rindler coordinates $(\tilde{t},\tilde{x})$ with a scaling factor $\lambda $, and is followed in a sequence by the special conformal transformation $K (\rho = 1/2\alpha)$, and a translation $T (-\alpha)$ defined in the usual way to place the diamond centered at the spacetime origin. 

After performing the transformation $M(\alpha;\lambda)$, the Minkowski coordinates $( t , x )$ of the diamond region $\mathrm{D}$ are the conformal image of the Minkowski coordinates $( \tilde{t} , \tilde{x} )$ of the right Rindler wedge, given by
 \begin{equation}
 \! \! \!  \! \! \!  \! \! \!  \! \! \!
(\tilde{t}, \tilde{x}) 
\overset{\displaystyle M(\alpha; \lambda) }{\xrightarrow{\hspace*{0.8cm}} }  
(t,x)
 : \; \; \; \;\; \; \; \;
     \frac{ t }{\alpha} 
     = \frac{ 2 \, \tilde{t}/\tilde{\alpha}
    }{ F_{+} ( \tilde{t}/\tilde{\alpha}, \tilde{t}/\tilde{\alpha}) }
        \; , \qquad  
   \frac{ x }{\alpha} 
  = - 
  \frac{
  N ( \tilde{t}/\tilde{\alpha}, \tilde{t}/\tilde{\alpha}) 
    }{ F_{+} ( \tilde{t}/\tilde{\alpha}, \tilde{t}/\tilde{\alpha}) }
    \label{eq:App_rindler-diamond_MR_R-to-D} 
  \; ,
\end{equation}
with its inverse mapping 
 $	M^{-1}(\alpha; \lambda)  =
 \Lambda^{-1} (\lambda) \circ K^{-1}(1/2\alpha)\circ T^{-1}(-\alpha) 
 = \Lambda (1/\lambda) \circ K (-1/2\alpha) \circ T(\alpha) $
being
\begin{equation}
 \! \! \!  \! \! \!  \! \! \!  \! \! \!
  (t,x)
\overset{\displaystyle M^{-1}(\alpha; \lambda) }{\xrightarrow{\hspace*{0.8cm}} } 
 (\tilde{t}, \tilde{x}) 
 : \; \; \; \;\; \; \; \;
    \frac{ \tilde{ t } }{\tilde{ \alpha} } 
     = \frac{2  \,  t/\alpha
   }{ F_{-} ( t/ \alpha, t/\alpha )    }
     \; , \qquad\qquad
      \frac{ \tilde{ x } }{\tilde{ \alpha} }     = 
    \frac{ N ( t/ \alpha, t/\alpha )   }{ F_{-} ( t/ \alpha, t/\alpha )    }
  \label{eq:App_rindler-diamond_MR_D-to-R}
    \; ,
 \end{equation} 
where, for both types of spacetime coordinates, $(t,x)$  and $( \tilde{t} , \tilde{x} )$,
the functions
\begin{equation}
 F_{\pm} ( t/ \gamma, x/\gamma )     
 =
 \left( x/\gamma \pm 1 \right)^2
 -
 \left( t/\gamma \right)^2
\; \; \;  \; \; \text{and} \; \; \; \; \; 
 N ( t/ \gamma, x/\gamma ) 
=
    1 -  ( x/\gamma )^2  +  ( t/\gamma )^2 
 \; ,
 \label{eq:App_rindler-diamond_MR_F-N-functions}
\end{equation}
are defined. In Eq.~(\ref{eq:App_rindler-diamond_MR_F-N-functions}),
$\gamma = \alpha, \tilde{\alpha}$ are the basic
scale units in diamond and Rindler spacetimes respectively; and the latter is given by
 \begin{equation}
 \tilde{\alpha } =
 \frac{2 \alpha}{\lambda}
 \; .
 \label{eq:App_tilde-alpha_def} 
\end{equation}
The definition of the basic functions 
$ F_{\pm} ( t/ \gamma, x/\gamma )  $    
and 
$ N ( t/ \gamma, x/\gamma ) $
of Eq.~(\ref{eq:App_rindler-diamond_MR_F-N-functions})
is due to the form of the special conformal transformation~(\ref{eq:sct_MR}).
In particular, for the transformation $\tilde{x}^{\mu} \longrightarrow x^{\mu}$,
the numerator in Eq.~(\ref{eq:sct_MR}) 
is $ F_{+} ( t'/ \gamma, x'/\gamma )  $, with $\gamma = 2 \alpha$; the inverse transformation would define 
a denominator $ F_{-} ( t''/ \gamma, x''/\gamma )  $,
; and $ N ( t'/ \gamma, x'/\gamma ) $ corresponds to the $x$-component 
($\mu =1 $) in both cases.
However, because of the dilatation factor $\lambda$, when  $x^{ \prime \mu} = \lambda \tilde{x}^{\mu}$,
for the transformation $\tilde{x}^{\mu} \longrightarrow x^{\mu}$,
these functions, when written in terms of the original Minkowskian coordinates of Rindler spacetime,
 take the form 
\begin{equation}
 \tilde{F}_{+} \equiv F_{+} ( \tilde{t}/ \tilde{\alpha}, \tilde{x}/\tilde{\alpha} )  
\; \;  \; , \; \; \;
 \tilde{N} \equiv N ( \tilde{t}/ \tilde{\alpha}, \tilde{x}/\tilde{\alpha} )  
 \; ,
 \label{eq:App_tilde-functions}
 \end{equation} 
where $\tilde{\alpha}= \gamma/\lambda = 2 \alpha/\lambda$.
As a result, after performing a final translation $T(-\alpha)$ to place the diamond centered at the spacetime origin, 
Eqs.~(\ref{eq:rindler-diamond_MR_R-to-D}) and (\ref{eq:App_rindler-diamond_MR_R-to-D})
are established. 
A similar analysis leads to the functions 
${F}_{-} \equiv F_{-} ( t/ \alpha, x/\alpha )$ and $N \equiv N ( t/ \alpha, x/\alpha )$ 
for the inverse transformation 
$ x^{\mu}  \longrightarrow  \tilde{x}^{\mu}$, thus establishing 
Eqs.~(\ref{eq:rindler-diamond_MR_D-to-R}) and (\ref{eq:App_rindler-diamond_MR_D-to-R}).

Finally, using the light-cone variables (\ref{eq:light-cone-coordinates}),
the transitional transformation equations~(\ref{eq:App_rindler-diamond_MR_R-to-D})--(\ref{eq:App_tilde-alpha_def})
 can be rewritten with
 \begin{equation}
 F_{\pm} ( t/ \gamma, x/\gamma )     
 =
 \left( 1 +  U_{\pm}/\gamma \right)  \left( 1 -  U_{\mp}/\gamma \right)
\; \; \;  \; \; \text{and} \; \; \; \; \; 
 N ( t/ \gamma, x/\gamma ) 
=
 1 + \left( U/\gamma \right) \left( V/\gamma \right)
 \; ,
 \label{eq:App_rindler-diamond_MR_F-N-functions_light-cone}
\end{equation}
 where $\gamma = \alpha, \tilde{\alpha}$ for $F_{\mp}$; and the corresponding 
 types of spacetime coordinates,   $(t,x)$  and $( \tilde{t} , \tilde{x} )$,
along with their associated light-cone variables 
$ U_{\pm}$ and $ \tilde{U}_{\pm}$ should be used.
In particular, for the transformation $ x^{\mu}  \longrightarrow  \tilde{x}^{\mu}$, 
the generic Eqs.~(\ref{eq:App_rindler-diamond_MR_F-N-functions_light-cone})
lead to
  \begin{equation}
 \! \! \!  \! \! \!  \! \! \!  \! \! \!
     \frac{ \tilde{t} }{\tilde{\alpha}} 
     = \frac{ U/\alpha + V/\alpha 
    }{ \left( 1 +  U/\alpha \right) \left( 1 - V/\alpha \right) }
        \; , \qquad  
   \frac{ \tilde{x} }{\tilde{\alpha}} 
  =
  \frac{
1 + \left( U/\alpha \right) \left( V/\alpha \right)
    }{ \left( 1 +  U/\alpha \right) \left( 1 - V/\alpha \right) }
    \label{eq:App_rindler-diamond_MR_R-to-D_light-cone}
  \; ,
\end{equation}
 whence the Rindler light-cone variables $\tilde{U}_{\sigma}^{(\epsilon)} = \epsilon ( \tilde{t} + \sigma \tilde{x} ) $
take values according to 
  \begin{equation}
\frac{  \tilde{U}_{\sigma}^{(\epsilon)} }{\tilde{\alpha}} 
= \sigma 
\left(
\frac{ 1 + {U}_{\sigma}^{(\epsilon)}/\alpha
}{  1 - {U}_{\sigma}^{(\epsilon)}/\alpha } 
\right)^{\sigma}
 \; .
  \label{eq:App_rindler-diamond_Mink-lightcone_R-to-D_gen}
  \end{equation}
Then, the relation~(\ref{eq:App_rindler-diamond_Mink-lightcone_R-to-D_gen}) directly gives
Eq.~(\ref{eq:rindler-diamond_Mink-lightcone_R-to-D}) with the assignments $\sigma = \pm 1$.

\subsection{Second step: Diamond (Rindler-like) coordinates}

 In the second step, the diamond coordinates $ \eta, \xi \in (-\infty,\infty)$ (and $\epsilon = \pm 1$ for $\mathrm{D}$ and  $\overline{\mathrm D}$,  respectively) are generally defined via
  \begin{equation}
 \tilde{t}
    = 
        \kappa    \, \epsilon
     \frac{1}{a} \,
\,e^{a \xi }
    \sinh (a \eta )
    \; ,
    \qquad \qquad 
       \tilde{x}
   = 
   \kappa     \, \epsilon
   \frac{1}{a} 
   \,e^{a \xi }
    \cosh (a \eta )
    \; ,
    \label{eq:App_Rindler-right_Minkowski_general}
\end{equation}
where $a$ is a standard Rindler acceleration parameter and $\kappa$ is an arbitrary scale factor.
In this analysis, it should be noted that
the sign convention in Eqs.~(\ref{eq:Rindler-right_Minkowski}) and (\ref{eq:App_Rindler-right_Minkowski_general})
enforces the physical condition that the Rindler Killing vector 
$\partial_{\eta} \propto \tilde{x} \partial_{\tilde{t}} + \tilde{t}\partial_{\tilde{x}}$
 (boost in the $\tilde{x}$ direction), 
which is timelike in the Rindler wedges $\mathrm{L}$
and $\mathrm{R}$, is future directed in $\mathrm{R}$, but past-directed 
in region $\mathrm{L}$. 
Thus, $ \partial_{\eta}$ is opposite $ \partial_{\tilde{t}}$ in the Rindler wedge $\mathrm{L}$~\cite{Olson-Ralph:2011}.
This could be more straightforwardly described by a coordinate $\rho \in (-\infty,\infty)$ 
replacing $ \epsilon\, e^{a \xi } /a  $ above~\cite{Takagi:1986},
but the use of $\xi$ simplifies the formulas using light-cone coordinates.
Moreover,
for the wedges $\mathrm{F}$ and $\mathrm{P}$, the Killing vector 
 $\partial_{\eta}$ is spacelike, with a reversal of the roles of spatial and temporal coordinates.
Finally, these standard assignments of Rindler spacetime are conformally mapped into diamond spacetime via
Eq.~(\ref{eq:App_rindler-diamond_MR_R-to-D}).

 In addition, Eq.~(\ref{eq:App_Rindler-right_Minkowski_general}) implies that
 \begin{equation}
\frac{  \tilde{U}_{\sigma}^{(\epsilon)} }{\tilde{\alpha}} 
=\delta  \sigma \epsilon 
\, \exp \left[  \sigma \epsilon   a {u}_{\sigma}^{(\epsilon)} \right]
 \; ,
 \end{equation}
 where 
 \begin{equation}
 \delta = \kappa \lambda/(2 a \alpha)
 \end{equation}
  is dimensionless. 
 Combining these equations,
 the relation between light-cone diamond and Minkowski variables follows,
 \begin{equation}
\delta^{\sigma} \epsilon 
\, \exp \left[ \epsilon a {u}_{\sigma}^{(\epsilon)} \right]
=
\frac{ 1 + {U}_{\sigma}^{(\epsilon)}/\alpha 
}{  1 - {U}_{\sigma}^{(\epsilon)}/\alpha } 
 \; .
 \label{eq:App_light-cone_diamond-Rindler-relations_gen}
 \end{equation}
In particular, for $\epsilon = +$ (interior diamond region D),
 \begin{equation}
\delta \, e^{av} \,
=
\frac{ 1 + V/\alpha 
}{  1 - V/\alpha  } 
 \; \; \; \; \text{and} 
 \; \; \; \;
\delta^{-1}  e^{au} \,
=
\frac{ 1 + U/\alpha 
}{  1 - U/\alpha  } 
 \; 
 \label{eq:App_light-cone_diamond-Rindler-relations_D}
 \end{equation}
(and similar relations for $\epsilon = -1$, with sign changes and inversions).

\subsection{Constraints on parameters and physical scales}

In this section,
the analysis of the general mapping between Rindler spacetime and diamond spacetime
 has not yet enforced any specific constraints related to physical scales. 
Clearly, this is an important requirement to establish the final equations.
  Now, Eq.~(\ref{eq:App_Rindler-right_Minkowski_general}) includes 
 the scale factor $\kappa$ that is needed to adjust the physical dimensions in diamond space to be compatible with
those in Rindler space. 
In addition, there has to be a specific relation between the diamond half-size $\alpha$ and the acceleration parameter $a$.
In particular, the value of diamond temperature and associated field-theory consequences 
(including quantum effects such as entanglement) dependent on these adjustments, which reduce the equations to the simple form~(\ref{eq:Rindler-right_Minkowski}) of Sec.~\ref{sec:diamond-geometry}.

The correct choice of constraints can be justified in a number of ways. For example, expand the relations in Eq.~(\ref{eq:App_light-cone_diamond-Rindler-relations_D}) near the center of the diamond, 
and enforce their mutual compatibility. With the light-cone variables being zero at the center of the diamond,
Eq.~(\ref{eq:App_light-cone_diamond-Rindler-relations_D}) gives
 $\delta (1 + av) \sim 1 + 2 V/\alpha$ and $\delta^{-1} (1 + au) \sim 1 + 2 U/\alpha$.
 These relations imply that $\delta =  1$ and  $a = 2/\alpha$, which combined with
 $\delta = \kappa \lambda/(2 a \alpha)$, 
 lead to the constraints 
 \begin{equation}
\kappa = \frac{4}{\lambda} \Leftrightarrow \delta = 1
  \; \; \; \; \text{and} 
 \; \; \; \;
   a = \frac{2}{\alpha}
\; .
 \label{eq:light-cone_diamond-Rindler-constraints}
 \end{equation}
These constraints
precisely fix the value of the acceleration parameter in relation to $\alpha$ and enforce the appropriate scale relations 
(between $\lambda $ and $\mu$) that guarantee the correct physical dimensions of the transformed spacetimes. 
As a result, Eq.~(\ref{eq:light-cone_diamond-Rindler-constraints})
reduces the formulas of this appendix to those of Sec.~\ref{sec:diamond-geometry}---see next subsection. 

Finally, one can verify the consistency of the  appropriate scale relations, $ \kappa = 4/\lambda$
by examining the metric.
First, the complete composite mapping can be represented 
by extending~(\ref{eq:mapping_diamond-Rindler-Minkowski}), 
with the labeling of the intermediate variables:
\begin{equation}
\; (\eta, \xi) \;
%
\substack{ \text{Rindler} \\ \text{mapping} \\ 
  \xrightarrow{\hspace*{0.45in}} \\   \text{Eq.~} (\ref{eq:App_Rindler-right_Minkowski_general})  \\ \; }
%
%
\; (\tilde{t},\tilde{x}) \;
%
\substack{  \\ \text{dilatation} \\   \xrightarrow{\hspace*{0.45in}} \\ \Lambda (\lambda)  }
\; (t',x') \;
%
\substack{ \text{special conformal} \\   \text{transf} \\ \xrightarrow{\hspace*{0.45in}} \\ 
 K(1/2\alpha) \\ \;  }
%
\; (t'',x'') \;
%
\substack{ \; \\ \text{translation}  \\ \; \xrightarrow{\hspace*{0.45in}} \\ T(-\alpha) \\ \; }
%
\; (t,x) \;
\label{eq:mapping_diamond-Rindler-Minkowski_2}
\end{equation}
  Then, from the relation between the following two metrics
\begin{equation}
 ds^2
 = - dt^2 + dx^2
 \; \; \; \; \text{and} 
 \; \; \; \;
 d\tilde{s}_{R}^2=
  \kappa^2 \, 
   \underbrace{
  e^{2a \xi}
  \left( -d \eta^2 + d \xi^2 \right)}_{\displaystyle ds_{R}^2}
  \; ,
\end{equation} 
i.e.,
the standard Minkowskian metric $ds^2= g_{\mu \nu} dx^{\mu} dx^{\nu}$  [with metric elements $\text{diag} (-1, 1)$]
and the rescaled Rindler-diamond metric $d\tilde{s}_{R}^2$, 
where the usual scaling for $d {s}_{R}^2$ corresponds to $\kappa =1$.
The metric is transformed in the implementation of the mapping~(\ref{eq:mapping_diamond-Rindler-Minkowski_2}), 
where two nontrivial scalings are involved: a simple scaling dilatation with factor $\lambda$
and a conformal transformation with conformal scaling factor $\Omega$,
so that
\begin{equation}
 ds^2
 =
\left(  \frac{\lambda}{\Omega} \right)^2 \, d\tilde{s}_{R}^2
=
\left(  \frac{\lambda \kappa }{\tilde{F}_{+}} \right)^2 \, ds_{R}^2
 \; .
 \label{eq:metric}
\end{equation}
Moreover, the conformal scaling factor is given by
\begin{equation}
\Omega \equiv
 \tilde{F}_{+} 
= F_{+} (\tilde{t}/\tilde{\alpha}, \tilde{x}/\tilde{\alpha})
=
 \left( 1 - \tilde{U}/\tilde{\alpha} \right)
 \,  \left( 1 + \tilde{V}/\tilde{\alpha} \right)
 =
4  \left( 1 + U/\alpha \right)^{-1}
 \,  \left( 1 - V/\alpha \right)^{-1}
  \; ,
 \label{eq:conformal-scaling}
\end{equation}
where the ``tilde functions,''
 including $\tilde{F}_{+} $, are given
in Eq.~(\ref{eq:App_tilde-functions}). 
The rightmost side of Eq.~(\ref{eq:conformal-scaling}) shows 
the approximate value near the center of the diamond
(with diamond coordinates $\eta, \xi \approx 0$) and is $F_{+} = 4$. Therefore, 
the Rindler and Minkowski-diamond scales can be assessed via their spacetime
variables $(t,x)$ and $(\eta, \xi)$ (which, in flat spacetime, correspond to specific lengths), leading to
the scaling constraint $\lambda \kappa/4 = 1$ as in Eq.~(\ref{eq:light-cone_diamond-Rindler-constraints}). 
 
 \subsection{Final expressions for the coordinate transformations}

 With the constraints imposed by the physical scaling conditions, the final expressions for the coordinate transformations take a somewhat simpler form.
 
Specifically, the constraints of Eq.~(\ref{eq:light-cone_diamond-Rindler-constraints})
reduce the formulas involving diamond coordinates $(\eta, \xi)$, i.e.,
Eqs.~(\ref{eq:App_Rindler-right_Minkowski_general})--(\ref{eq:App_light-cone_diamond-Rindler-relations_D})
to those of Sec.~\ref{sec:diamond-geometry}. 
In particular,
Eq.~(\ref{eq:Rindler-right_Minkowski}) is established for the setup of $(\eta, \xi)$
with $\kappa/a = \tilde{\alpha}$.

Most importantly, this also implies that
 {\em the composite mapping $(t,x) \longrightarrow (\eta, \xi)$,
and its inverse, is
 independent of the scaling factor $\lambda$---this is unlike the transitional $\lambda$-dependent 
 mapping~(\ref{eq:rindler-diamond_MR_R-to-D}).\/}
Specifically, this
procedure has to be set up separately for the four different regions of Rindler spacetime with separate coordinate patches.
For the most relevant diamond
regions $\mathrm{D}$ and $\overline{\mathrm D}$ ($\epsilon = \pm$), the final expressions take the form
  \begin{equation}
\frac{ 2 \eta }{\lambda} 
  =
  \tanh^{-1}
 \left[
 \frac{ 2 \, t/ \alpha}{
 N(t/\alpha; x/\alpha)
     }
 \right]
\; \; \; , \; \; \;
\frac{ 2 \xi }{\lambda} 
  =
 \ln
  \left(
  {\frac{
\sqrt{ \left[ N(t/\alpha; x/\alpha) \right]^2 - (2t/\alpha)^2 }
    }{
F_{-}(t/\alpha; x/\alpha)
  }
  } 
  \right)
 \; ;
 \label{eq:diamond-coordinates_MR}
 \end{equation}
 it should be noticed that one could use
 a different notation 
$(\bar{\eta}, \bar{\xi})$ for  $\overline{\mathrm D}$ ($\epsilon = -$), as in 
Eq.~(\ref{eq:rindler-diamond_lightcone1_R-to-D_ext}).
In addition,  for the two regions in 
 $\overline{\overline{\mathrm D}}$,
 the reversal $\tilde{t} \leftrightarrow \tilde{x}$ can be enforced, 
yielding the switch of variables   $2t/\alpha  \leftrightarrow  N(t/\alpha; x/\alpha) $
in Eq.~(\ref{eq:diamond-coordinates_MR}), with sets of different Rindler parameters.
 We have thus covered with diamond coordinates all regions of the maximally extended diamond spacetime in a scale-independent manner.

Finally, the corresponding expressions
between light-cone diamond and Minkowski variables
can be derived by recasting Eq.~(\ref{eq:App_light-cone_diamond-Rindler-relations_gen}) 
 into the form
  \begin{equation}
 \epsilon 
\, \exp \left[2 \epsilon {u}_{\sigma}^{(\epsilon)}/\alpha \right]
=
\frac{ 1 + {U}_{\sigma}^{(\epsilon)}/ \alpha 
}{  1 - {U}_{\sigma}^{(\epsilon)}/\alpha } 
 \; ,
 \label{eq:master-light-cone_relations}
 \end{equation}
 which are, of course, also $\lambda$ independent. 
 Equation~(\ref{eq:master-light-cone_relations})
gives the explicit  Eqs.~(\ref{eq:rindler-diamond_lightcone1_R-to-D_int})
and (\ref{eq:rindler-diamond_lightcone1_R-to-D_ext})
 for $\epsilon = \pm 1$, respectively.

\section{Diamond Geometry---Global Properties and Classification of Causal Regions}
 \label{sec:diamond-geometry_global}
 
 It is noteworthy that, in the existing literature, 
only the transformation of the right Rindler wedge $\mathrm{R}$ to the main diamond $\mathrm{D}$ is fully displayed.
However, Eqs.~(\ref{eq:rindler-diamond_MR_R-to-D}) and (\ref{eq:rindler-diamond_MR_D-to-R}) 
cover all of Minkowski space, allowing for a complete mapping of all the wedges 
and identifying the properties of both the interior and the exterior of the causal diamond.
 In particular, these equations display the setup of the diamond geometry, as follows:
 \begin{itemize}
 \item
 The straight-line boundaries $\tilde{x} = \pm \tilde{t}$ 
 of the Rindler wedges transform to the four straight lines that limit both the diamond itself and the demarcations of the 
 boundaries of its external region:
 $t = \pm ( x \pm \alpha )$ (with all four combinations of signs).
 \item
The four vertices of the diamond can be identified from the relevant origin and the different types of infinity of Rindler space:
 the left vertex  $(t, x)=(0, -\alpha)$ from the Rindler-space origin;
 the right vertex  $(t, x)=(0, \alpha)$ from the Rindler-space spatial asymptotic infinity 
 (along the axis $\tilde{x}$) ;
  the top vertex  $(t, x)=(\alpha,0)$ from the Rindler-space future asymptotic infinity of accelerated observers;
  and the bottom vertex  $(t, x)=-(\alpha,0)$ from the Rindler-space past asymptotic infinity of accelerated observers.
 \item
 All of the above assignments, as well as more detailed information, can be easily derived using the relationship
 between the 
 light-cone variables $(\tilde{V},\tilde{U}) \longleftrightarrow (V,U)$, as given by 
 Eq.~(\ref{eq:rindler-diamond_Mink-lightcone_R-to-D}) and/or their inverse relations.
 
 In particular, 
 (i) the signs of the coordinates $(\tilde{V},\tilde{U})$ permit the identification of the Rindler wedges;
 and (ii) the values of $|V|$ and $ |U|$,  compared with unity,
 permit the identification of the location
 of the correspondingly conformally mapped regions in diamond spacetime.
  The result of this full mapping is shown in Fig.~\ref{fig:wedges}, where 
one can identify the following: 
\begin{itemize}
\item
The known right Rindler wedge 
($\tilde{V}>0, \tilde{U}<0$)
mapping  into the causal diamond 
($|V|,|U|<1$),
as described above, and in accordance with the literature.
\item
The left Rindler wedge 
($\tilde{V}<0, \tilde{U}>0$)
mapping into four unconnected diamond wedges attached to the four vertices
($|U|,|V|>1$). 

\item 
The future 
($\tilde{V}>0, \tilde{U}>0$)
and past Rindler wedge 
($\tilde{V}<0, \tilde{U}<0$)
mappings into two unconnected semi-infinite rectangular regions each, which 
extend to the outside of the causal diamond boundaries.
(The image of the future wedge consists of the top left and bottom right rectangles with $|V|< 1,|U|>1$. And
the image of the past wedge 
consists of the top right and bottom left rectangles with  $|V|> 1,|U|<1$.)

\end{itemize}
 \end{itemize}

\section{Canonical Quantization and Bogoliubov coefficients in causal diamonds}
\label{sec:appCanonQuant_BogCoefs}
In this appendix, we review the canonical quantization in diamond coordinates and establish a comparison between two quantization schemes, in terms of diamond and Minkowski modes by computing the Bogoliubov coefficients.

\subsection{Canonical Quantization}

From Eqs.~(\ref{eq:K-G-eq_Minkowski}) and (\ref{eq:K-G-modes_Minkowski}), the quantum field can be expanded in Minkowski modes as
\begin{equation}
\Phi 
= \Phi_{+} (V)+ \Phi_{-} (U)
=
\int_{0}^{\infty}
 dk \,
\bigl[
a_{+,k} \, f_{+,k}(V)
+ a_{-,k} \, f_{-,k}(U)
+ \mathrm{H.c.}
\bigr]
\; ,
\label{eq:field-expansion_Minkowski}
\end{equation}
where  H.c.\ stands for the Hermitian conjugate, and it is understood that the field $\Phi (x,t)$ is written in terms of Minkowski null coordinates.
In Eq.~(\ref{eq:field-expansion_Minkowski}), the field operators satisfy the commutator relations
\begin{equation}
[a_{\sigma,k} , a^{\dagger}_{\sigma',k'} ]
= \delta_{\sigma, \sigma'} 
\,
\delta (k-k')
\; , \; \; \;
[a_{\sigma,k} , a_{\sigma',k'} ] = 0
\; , \; \; \;
[a^{\dagger}_{\sigma,k} , a^{\dagger}_{\sigma',k'} ]
= 0
\; 
\label{eq:field-operator-algebra_Minkowski}
\end{equation}
for all traveling directions $\sigma,\sigma'$ and all Minkowski frequencies $k,k'$. Here, the left and right movers, $ \Phi_{+} (V)$ and $ \Phi_{-} (U)$,
do not interact with each other, i.e., all the left-mover operators $a_{+,k} $ commute with all the right-mover operators $a_{-,k'} $ for all Minkowski frequencies $k,k'$; thus, they can be treated independently. Then, the Minkowski vacuum $\left| 0 \right\rangle^{\!\mathcal M}$ is the state is that satisfies 
\begin{equation}
a_{\pm,k} \, 
\left| 0 \right\rangle^{\!\mathcal M}
= 0
\; ,
\label{eq:Minkowski-vacuum}
\end{equation}
for all frequencies $k$. The notation ${\mathcal M}$ is used for the basis defined by the set of inertial modes~(\ref{eq:K-G-modes_Minkowski}) in Minkowski spacetime.

An alternative quantization scheme can be set up in diamond coordinates, for which the solutions inside and outside the diamond form a complete set. Thus,
the field can be expanded in the form
\begin{equation}
\Phi
 =\Phi_{+}(v(V))+\Phi_{-}(u(U)) 
\; ,
\label{eq:field-expansion_diamond}
\end{equation}
where $\Phi_{+}(v(V))$ and $\Phi_{-}(u(U))$ are the same operators
$ \Phi_{+} (V)$ and $ \Phi_{-} (U)$ as in Eq.~(\ref{eq:field-expansion_Minkowski})
but rewritten in terms of the diamond 
modes~(\ref{eq:K-G-modes-generic_diamond})--(\ref{eq:ext-minus-mode}), i.e., 
\begin{equation}
\Phi_{+}
=
\int_{0}^{\infty} d\omega 
\bigl[
{b}_{+, \omega}^{(\mathrm{int})}
g_{+, \omega}^{(\mathrm{int})}(V)
+{b}_{+, \omega}^{(\mathrm{ext})}
g_{+, \omega}^{(\mathrm{ext})}(V)
+\mathrm{H.c}
\bigr]
\label{eq:field-expansion_diamond_left-movers}
\end{equation}
 and 
\begin{equation}
\Phi_{-}
=
\int_{0}^{\infty} d\omega 
\bigl[
{b}_{-, \omega}^{(\mathrm{int})}
g_{-, \omega}^{(\mathrm{int})}(U)
+{b}_{-, \omega}^{(\mathrm{ext})}
g_{-, \omega}^{(\mathrm{ext})}(U)
+\mathrm{H.c}
\bigr]
\; .
\label{eq:field-expansion_diamond_right-movers}
\end{equation}
In Eqs.~(\ref{eq:field-expansion_diamond})-(\ref{eq:field-expansion_diamond_right-movers}),
the field operators are subject to the commutator relations
\begin{equation}
[b^{(\mathrm{int})}_{\sigma,\omega} , b^{(\mathrm{int}) \dagger}_{\sigma',\omega'} ]
= \delta_{\sigma, \sigma'} 
\,
\delta (\omega-\omega')
\; , \; \; \;
[b^{(\mathrm{ext})}_{\sigma,\omega} , b^{(\mathrm{ext}) \dagger}_{\sigma',\omega'} ]
= \delta_{\sigma, \sigma'} 
\,
\delta (\omega-\omega')
\; ,
\end{equation}
along with all the other commutators being identically zero. Again, these relations involve all traveling  directions $\sigma,\sigma'$
and diamond frequencies $\omega, \omega'$. Then, the diamond vacuum state $\left| 0 \right\rangle^{\!\mathcal D} $
satisfies 
\begin{equation}
b^{(\mathrm{int})}_{\pm,\omega} \, 
\left| 0 \right\rangle^{\!\mathcal D} = 0
\; \; 
\text{and}
\; \;
b^{(\mathrm{ext})}_{\pm,\omega} \, 
\left| 0 \right\rangle^{\!\mathcal D}  = 0
\label{eq:diamond-vacuum}
\end{equation}
for all diamond frequencies $\omega$. 
This is the conformally mapped counterpart of the Rindler vacuum, via Eqs.~(\ref{eq:App_rindler-diamond_MR_R-to-D})--(\ref{eq:App_rindler-diamond_MR_F-N-functions}), and the notation ${\mathcal D} $ stands for the basis defined by the set of diamond modes~(\ref{eq:K-G-modes-generic_diamond}).

{\em Both the remarkable thermal behavior of diamond observers and the ensuing entanglement degradation 
are a consequence of the inequivalence of these quantization schemes in Minkowski and diamond coordinates.\/}
 This issue is discussed next.

\subsection{Bogoliubov coefficients}

The comparison between the two quantization schemes, in terms of diamond and Minkowski modes, can be established through the Bogoliubov transformations
\begin{equation} 
g_{\sigma, \omega}^{(\mathrm{int})} (U_{\sigma} )
=
\int_0^\infty 
dk
\bigl[
\alpha_{\sigma, \omega k}^{(\mathrm{int})}
\, f_{\sigma, k} (U_{\sigma} )
+
\beta_{\sigma, \omega k}^{(\mathrm{int})}
\, f^*_{\sigma, k} (U_{\sigma} )
\bigr]
 \; 
\label{eq:Bogoliubov-int_expansion}
\end{equation}
and
\begin{equation}
g_{\sigma, \omega}^{(\mathrm{ext})} (U_{\sigma} )
=
\int_0^\infty 
dk
\bigl[
\alpha_{\sigma, \omega k}^{(\mathrm{ext})}
\, f_{\sigma,k} (U_{\sigma} )
+ \beta_{\sigma, \omega k}^{(\mathrm{ext})}
\, f^*_{\sigma,k} (U_{\sigma} )
\bigr]
 \; .
\label{eq:Bogoliubov-ext_expansion}
\end{equation}
where $\sigma = \pm$ for the left- and right-movers, with $\alpha_{\sigma, \omega k}^{(\mathrm{int})}$, $\beta_{\sigma, \omega k}^{(\mathrm{int})}$,  $\alpha_{\sigma, \omega k}^{(\mathrm{ext})}$, and $\beta_{\sigma, \omega k}^{(\mathrm{ext})}$ being the Bogoliubov coefficients. 
The expressions for the diamond modes~(\ref{eq:int-plus-mode})-(\ref{eq:int-minus-mode}), along with the corresponding Minkowski modes~(\ref{eq:K-G-modes_Minkowski}), show that not only the left- and right-movers are independent, but their associated equations have the same form, so that their Bogoliubov coefficients are identical:
 $\alpha_{+, \omega k}^{(\mathrm{int})}
 =
 \alpha_{-, \omega k}^{(\mathrm{int})}$
 and
  $\beta_{+, \omega k}^{(\mathrm{int})}
 =
 \beta_{-, \omega k}^{(\mathrm{int})}$,
and similarly with the external diamond modes~(\ref{eq:ext-plus-mode})-(\ref{eq:ext-minus-mode}). In addition, consistent with the invariant nature of the field, the creation and annihilation operators satisfy the reciprocal Bogoliubov relations (covariant rather than contravariant transformations),
\begin{equation} 
{b}_{\sigma, \omega}^{(\mathrm{int})}
=
\int_0^\infty 
dk
\bigl[
\alpha_{\sigma, \omega k}^{(\mathrm{int})*}
\, {a}_{\sigma, k}
-
\beta_{\sigma, \omega k}^{(\mathrm{int})*}
\, {a}^{\dagger}_{\sigma, k}
\bigr]
 \; ,
\label{eq:Bogoliubov-int_expansion_op}
\end{equation}
and
\begin{equation}
{b}_{\sigma, \omega}^{(\mathrm{ext})}
=
\int_0^\infty 
dk
\bigl[
\alpha_{\sigma, \omega k}^{(\mathrm{ext})*}
\, {a}_{\sigma, k}
-
\beta_{\sigma, \omega k}^{(\mathrm{ext})*}
\, {a}^{\dagger}_{\sigma, k}
\bigr]
 \; .
\label{eq:Bogoliubov-ext_expansion_op}
\end{equation}

Then, the Bogoliubov coefficients can be computed by projection via the generic  Klein-Gordon inner product~\cite{Takagi:1986, Crispino:2008}
$(\Phi_1,\Phi_2) 
= 
 i \int_\Sigma 
 \Phi_1^* \stackrel{\leftrightarrow}{\partial^\mu} \Phi_2 
\, 
 \sqrt{-g}\;d\Sigma^\mu
$
(where the integral is performed on a spacelike hypersurface). However, a useful shortcut can be taken: given the specific form of the Minkowski modes~(\ref{eq:K-G-modes_Minkowski}), i.e., $f_{\sigma,k} (U_{\sigma} ) \propto  e^{-ik U_{\sigma} }$, it follows that  Eqs.~(\ref{eq:Bogoliubov-int_expansion}) and (\ref{eq:Bogoliubov-ext_expansion}) have the functional form of Fourier transforms with conjugate variables $k$ and $U_{\sigma} $.
Thus, it is straightforward to invert these transforms with respect to Minkowski frequencies $k$ in the positive and negative frequency ranges, to yield integral expressions for the Bogoliubov coefficients with respect to the light-cone coordinates ($U_{\sigma} = V,U$), 
\begin{equation}
\begin{aligned}
\alpha_{\sigma, \omega k}^{(\mathrm{int})}
& =
\sqrt{4 \pi k} 
\,
\int_{-\infty}^{\infty}
\frac{dU_{\sigma}}{2\pi} \,
g_{\sigma, \omega}^{(\mathrm{int})}(U_{\sigma}) \,
e^{ i kU_{\sigma}}
 =
\frac{\alpha}{2 \pi}
 \,
\sqrt{ \frac{\hat{k}}{\hat{\omega}} } 
\,
\int_{0}^{\infty}
d \hat{U}_{\sigma}  \,
\left( \frac{1+ \hat{U}_{\sigma}}{1-\hat{U}_{\sigma} } \right)^{- i \hat{\omega}/2}
 \,
e^{ i \hat{k} \hat{ U}_{\sigma} }
\\
& =
\frac{\alpha}{4} \,
\frac{  \sqrt{ \hat{\omega} \hat{k} } }{ 
 \sinh (\pi \hat{ \omega }/2 ) }
\, e^{ - i \hat{k} }
\, M(1-i \hat{\omega}/2 ,2, 2i \hat{k} )     
\; ,
\label{eq:Bogoliubov-int_alpha}
\end{aligned}
\end{equation}
and
\begin{equation}
\begin{aligned}
\beta_{\sigma, \omega k}^{(\mathrm{int})}
& =
\sqrt{4 \pi k} 
\,
\int_{-\infty}^{\infty}
\frac{dU_{\sigma}}{2\pi} \,
g_{\sigma, \omega}^{(\mathrm{int})}(U_{\sigma}) \,
e^{- i kU_{\sigma}}
=
\frac{\alpha}{2 \pi}
 \,
\sqrt{ \frac{\hat{k}}{\hat{\omega}} } 
\,
\int_{0}^{\infty}
d \hat{U}_{\sigma}  \,
\left( \frac{1+ \hat{U}_{\sigma}}{1-\hat{U}_{\sigma} } \right)^{- i \hat{\omega}/2}
 \,
e^{- i \hat{k} \hat{ U}_{\sigma} }
\\
& =
\frac{\alpha}{4} \,
\frac{  \sqrt{ \hat{\omega} \hat{k} } }{ 
 \sinh (\pi \hat{ \omega }/2 ) }
\, e^{ i \hat{k} }
\, M(1-i \hat{\omega}/2 ,2, -2i \hat{k} )   
\; ,
\label{eq:Bogoliubov-int_beta}
\end{aligned}
\end{equation}
where $M(\mu, \nu, z)$ is Kummer's confluent hypergeometric function, and all the variables with a hat are dimensionless rescaled versions with the diamond parameter $\alpha$ (see Appendix~\ref{sec:general-conformal-mappings}), i.e.,
 $\hat{\omega}=\omega\alpha$,
 $\hat{k}=k\alpha$,
 and $\hat{U}_{\sigma}=U_{\sigma}/\alpha$.
With appropriate adjustments in the definitions used, these results agree with Refs.~\cite{Su2016SpacetimeDiamonds, Foo2020GeneratingMirror}.
In particular, $\alpha_{\sigma, \omega k}^{(\mathrm{int})}$ can be obtained from $\beta_{\sigma, \omega k}^{(\mathrm{int})}$ by the replacement $k \rightarrow -k$, except within the prefactor $\sqrt{k}$. A similar calculation follows for the Bogoliubov coefficients relating the external modes with the Minkowski modes.
 
From the Bogoliubov coefficients, using standard technology~\cite{birrell-davies,Takagi:1986, Crispino:2008}, the particle number creation for each mode,
for  each propagation direction $\sigma$ and frequency $\omega$,  can be computed as 
$n_{\sigma,\omega} = 
\bra{0}
{b}_{\sigma, \omega}^{(\mathrm{int})\dagger} 
{b}_{\sigma, \omega}^{(\mathrm{int})}
\ket{0}
= 
 \sum_{k} \left| \beta_{\sigma, \omega k}^{(\mathrm{int})} \right|^2$.
In the continuum limit, this needs to be regularized and can be formally rewritten in the form 
$ 
 \int_{0}^{\infty}
 dk 
 \beta_{\sigma, \omega k}^{(\mathrm{int})*} 
 \beta_{\sigma, \omega' k}^{(\mathrm{int})}
  = \delta( \omega - \omega') n_{\sigma, \omega}$.
Specifically, this integral can be computed using the original integral expressions~(\ref{eq:Bogoliubov-int_beta}), showing~\cite{Su2016SpacetimeDiamonds}
that this yields a Bose-Einstein distribution $n_{\sigma, \omega}= (e^{\beta \omega} -1 )^{-1}$ for scalar fields, where the inverse temperature is $\beta = \pi \alpha$. A complete analysis involves writing the vacua (see Sec.~\ref{sec:Unruh-diamond-modes} in the main text). As a result, one concludes that the diamond vacuum is a thermal state with temperature given by Eq.~(\ref{eq:diamond-temperature}).

In addition to this standard analysis via Bogoliubov coefficients, this result on thermality has been shown by the use of an Unruh-DeWitt energy-scaled 
detector~\cite{Su2016SpacetimeDiamonds} and by the use of an open quantum systems approach~\cite{Chakraborty:2022oqs}.
In Sec.~\ref{sec:Unruh-diamond-modes}, this is further confirmed by modeling the Minkowski vacuum
with a generalization of Unruh's analytic continuation technique  (originally conceived for Rindler spacetime)~\cite{Unruh:1976}.

\end{document}